\documentclass{ws-ijmpa}
\usepackage[super,compress]{cite}
\usepackage{graphicx}

\usepackage{hyperref}
\usepackage{tensor}
\usepackage{bm}% Include figure files
\newcommand{\rr}{Ref.\ }
\newcommand{\rs}{Refs.\ }

\begin{document}
\markboth{Agam Shayit, S. A. Fulling, T. E. Settlemyre, Joseph Merritt}{Vacuum energy density and pressure inside a soft wall}

\catchline{}{}{}{}{}
%
%%%%%%%%%%%%%%%%%%%%%%%%%%%%%%%%%%%%%%%%%%%%%%%%%%%%%%%%%%%%%%%%%%%%

\title{Vacuum energy density and pressure inside a soft wall
}

\author{Agam Shayit}

\address{Department of Physics and Astronomy, Texas A\&M University, College Station, Texas 77843-4242, USA\\
Department of Computer Science and Engineering, Texas A\&M University, College Station, Texas 77843-3112, USA\\
agam@tamu.edu}

\author{S.~A.~Fulling\footnote{Corresponding author.}}

\address{Department of Physics and Astronomy, Texas A\&M University, College Station, Texas 77843-4242, USA\\
Department of Mathematics, Texas A\&M University, College Station, Texas 77843-3368, USA\\fulling@math.tamu.edu}

\author{T.~E.~Settlemyre}

\address{Department of Physics and Astronomy, Texas A\&M University, College Station, Texas 77843-4242, USA\\
Department of Mathematics, Texas A\&M University, College Station, Texas 77843-3368, USA\\tommy7410@tamu.edu}

\author{Joseph Merritt \footnote{Present address: Department of Physics, University of Washington, Seattle, Washington 98195-1560, USA}}

\address{Department of Physics and Astronomy, Texas A\&M University, College Station, Texas 77843-4242, USA\\
Department of Mathematics, Texas A\&M University, College Station, Texas 77843-3368, USA\\jm117@uw.edu}

\maketitle

\begin{abstract}
In the study of quantum vacuum energy and the Casimir effect, it is desirable to model the conductor by a potential of the form $V(z)=z^\alpha$.  This ``soft wall" model was proposed so as to avoid the violation of the principle of virtual work under
ultraviolet regularization that occurs for the standard Dirichlet wall. The model was formalized for a massless scalar field, and the expectation value of the stress tensor has been expressed in terms of the reduced Green function of the equation of motion. In the limit of interest, $\alpha \gg 1$, which approximates a Dirichlet wall, a closed-form expression for the reduced Green function cannot be found, so   piecewise approximations incorporating the perturbative and WKB expansions of the Green function, along with  interpolating splines in the region where neither expansion is valid, have been developed. After reviewing this program, in this article we apply the scheme to the  wall with $\alpha=6$ and use it to compute the renormalized energy density and pressure inside the cavity for various values of the conformal parameter. The consistency of the results is verified by comparison to their numerical counterparts and verification of the trace anomaly and the conservation law. Finally, we use the approximation scheme to reproduce the energy density inside the quadratic wall, which was previously calculated exactly but with some uncertainty. 

\end{abstract}

\ccode{PACS numbers: 03.70.+k,02.30.Mv,02.60.-x,11.10.Gh}

%\tableofcontents

\section{Introduction}

The Casimir effect in quantum field theory refers to the (usually attractive) force between the boundaries of a cavity in vacuum. This force is induced by the energy gradient resulting from the modification of the normal modes of the quantum field inside the cavity. The existence of this force entails that the expectation value of the energy density is nonzero. 
This force was first derived by Casimir in \rr\citen{originalCasimir} for an electromagnetic field and a perfect conductor. Many later works, such as \rs\citen{Fierz,Boyer,BenderHays}, simplify Casimir's model by replacing the EM field and the conductor with a scalar field and Dirichlet boundary conditions.

Such ``hard walls'' (perfectly reflecting boundaries) 
cannot be taken too literally as models of real physical systems, 
because they predict infinite deposits of energy nearby in the 
quantum fields.  A real conductor does not reflect waves of 
arbitrarily high frequency.  A common remedy, therefore, is to 
impose an ultraviolet cutoff in integrals over frequency.  This 
``regularization'' may be regarded in two ways.  First, it can be 
a temporary expedient to produce finite quantities that can be 
mathematically manipulated into a sum of terms that approach 
finite limits as the cutoff is removed and terms that diverge but 
have simple expressions in terms of local quantities (in the 
present case, describing the geometry of the boundary); the 
latter can safely be ignored, or absorbed into constants such as 
the mass of the wall (``renormalization'').  Alternatively, one 
might ask whether leaving the cutoff parameter finite would yield 
a qualitatively acceptable approximate model of the system 
originally envisaged.
	Exploration of this second path revealed \cite{refStar, estrada} that the 
finite cutoff theory violates the ``principle of virtual work'': 
the change in energy when a wall moves adiabatically is not equal 
to the work done by the pressure on the wall.  This observation 
(``pressure anomaly'') was ``covariantized'' by generalizing the 
frequency cutoff to regularization by point-splitting in an 
arbitrary direction.  From the point of view of renormalization, 
this allows the anomalous terms (which would be divergent anyhow) 
to be systematically discarded in the manner now standard in 
quantum field theory in curved space-time \cite{Christensen, wald1, wald2}. The finite-cutoff 
model, however, it leaves both ill-defined and seemingly
physically inconsistent.

A somewhat more realistic model of the boundary is a 
power-law potential (``soft wall'').  The root of the problem 
with virtual work is presumably that tinkering with the solution 
of the field theory after the fact has produced something that is 
not a solution of any self-consistent theory.  In contrast, 
adding an external potential (see Sec.~\ref{section:softWallModel}) creates a perfectly standard field theory, derivable from a Lagrangian, hence almost 
guaranteed to be consistent.  Indeed, this theory does not 
manifest a pressure anomaly.  It does require additional
regularization and renormalization to handle the interaction 
between the quantized field and the potential field in the bulk 
region, but the treatment of that problem is quite conventional.  
It is not claimed that this theory is a close model of anything 
existing in nature, but it shows behavior that is qualitatively 
what one would expect.  Compared to a full treatment of the 
interaction of the EM field with a realistic conductor (a major 
venture in condensed-matter theory), the model is comparatively 
tractable analytically and computationally with results that are 
easily understandable in terms of the asymptotics of solutions of 
differential equations.

This article reviews and completes a program
\cite{Dartmouth,HardAndSoftWalls,swout,interior,Symmetry} devoted to a single massless scalar field, $\phi$, interacting with a single soft wall. (Previous work by others was reviewed in the Appendix of \rr\citen{swout}.  Recent similar work on the electromagnetic field in inhomogeneous dielectric media includes \rs\citen{GL1,GL2,PMLDGFC,LMGKF,Zelnikov2021}). Most of the material not previously published is drawn from \rr\citen{Shayit}.

Sec.\  \ref{section:softWallModel} introduces the
model\cite{Dartmouth} and
presents the basic formulas for the normal modes, Green function,
and stress tensor of the quantum field
theory\cite{HardAndSoftWalls,swout,interior}.
Secs.\
\ref{section:perturbationTheory} and \ref{section:wkbApproximation} present the
perturbation theory and WKB approximation (developed in
\rs\citen{swout,interior}) used to approximate the
solutions at small and large values, respectively, of the
complexified wave number,~$\kappa$.  To obtain satisfactory
numerical results it was found necessary in \rr\citen{Shayit}
to extend these expansions systematically to arbitrarily high
order.  Then in Sec.\ \ref{section:theIntermediateRegime} we use these approximations to develop a piecewise approximation scheme for the Green function and its second derivative. We then compare the piecewise approximations to their numerical counterparts.

We use the analytical and numerical approximations to calculate the energy density and pressure inside the sextic soft wall. These results are plotted in Secs.~\ref{section:theEnergyDensityInside} and \ref{section:thePressureInside}. In Sec.~\ref{section:conservationLaws}, we verify that the approximated stress tensor upholds the trace identity and conservation law established in \rs\citen{interior,Symmetry}.

Finally, in Sec.~\ref{section:reproducingTheQuadratic}, we use the approximation scheme developed in Sec.~\ref{section:theIntermediateRegime} to reproduce the energy density inside the quadratic wall. We then compare the results to their analytical counterparts calculated in \rr\citen{interior}. 

Our conclusion, Sec.~\ref{section:conclusion}, 
includes a discussion of the prospects for eliminating in future 
work certain \emph{ad hoc} elements in the approximation scheme. It is followed by two appendices, the first on the numerical solution and the second on previously attempted approximation schemes.

\section{The Soft Wall Model}\label{section:softWallModel}

Understanding local energy density and pressure is
essential for general relativity.  It also clarifies the physics
of global energy and force calculations.\cite{hertzberg,rect}
Therefore, a prime objective is calculating expectation values of
the stress tensor, $\langle T^{\mu\nu}\rangle$.  Traditionally the
counterterms that arise in a regularized calculation of the
total energy have sometimes been thought to bear some relevance
to the actual energy of a conjectured ultraviolet-finite theory
to which the divergent idealized theory is an approximation.
Those terms, however, are extremely sensitive to the details of
the regularization, and any hope of taking them seriously in a
calculation with a standard exponential ultraviolet cutoff was
dashed by the discovery of an inconsistency between the
counterterms in the energy and the
pressure.\cite{refStar,estrada}  Namely, these divergent terms
(but not the finite remainders) violate the so-called principle
of virtual work\cite{B3}:  The pressure on a partition from one
side is
not equal to the negative of the derivative of the energy on that
side with respect to the position of the partition.
Whem the exponential ultraviolet cutoff is covariantized to
regularization by point-splitting, one realizes that the pressure
anomaly is, at root, an instance of the dependence of the
regularized stress on the direction of point separation,
thoroughly investigated by Christensen\cite{Christensen} in the
setting of a background gravitational field.  Separating the
points in a ``neutral'' direction (neither temporal, nor the
spatial direction of the pressure component concerned) yields
physically plausible results, but this ad hoc procedure is
clearly not logically satisfactory.
(The foregoing remarks are paraphrased from \rr\citen{tomsk}.)

Therefore, to obtain a physically acceptable model with a
finite cutoff it was proposed in \rs\citen{Dartmouth,HardAndSoftWalls} to replace the Dirichlet wall with the ``softer" counterpart
\begin{equation}\label{IntroSoftWallPotential}
    V(z) = 
    \begin{cases}
    0 & \text{if } z \leq 0,\\
    z^\alpha & \text{if } z > 0.
    \end{cases}
\end{equation}
Here $\alpha>0$ is a stiffness parameter,
which we ordinarily take to be an integer to avoid extraneous singular behavior at the origin. This potential interacts with a massless Klein--Gordon field according to Eq.\ (1.2) of \rr\citen{interior}, namely
\begin{equation}\label{MethodsGeneralizedKGLagrangian}
    \mathcal{L} = -\frac{1}{2}\partial_\mu \phi \partial^\mu \phi - \frac{V}{2}\phi^2.
\end{equation}

This Lagrangian density yields the generalized Klein--Gordon equation
\begin{equation} \label{IntroFieldeq}
-\frac{\partial^2\phi}{\partial t^2} + \nabla^2 \phi
= V(z) \phi,
\end{equation}
where without loss of generality one chooses units where $c=1$ and $\hbar=1$, so that both
$t$ and $z$ have units of inverse mass.
Indeed, $V$ takes the place of the mass squared in the usual Klein--Gordon equation.
As it stands this appears inconsistent with (\ref{IntroSoftWallPotential}), 
so, following \rr\citen{Dartmouth}, we momentarily adopt a maximally cautious notation that makes all the free parameters explicit and their dimensions transparent:
\begin{equation}\label{IntroSP}
V(z) = \lambda_0\left(\frac{z}{z_0}\right)^\alpha \qquad \text{if } z>0,
\end{equation}
where $z_0$ is a length and $\lambda_0$ is the square of a mass.
This parametrization, however, is obviously redundant, because of the special scaling behavior of the power function.
The strength of the potential is determined by a single coupling constant, $\hat{\lambda}=\lambda_0 z_0^{-\alpha}$,
or, alternatively, a single length scale,
$$\hat{z} = \left(\frac{z_0^\alpha}{\lambda_0}\right)^{\frac{1}{\alpha+2}}.$$
(Since $\alpha$ is assumed to be positive, the singularity at $\alpha=-2$ is no cause for concern.)
Then 
\begin{equation}\label{IntroPotentials} V(z) = \hat{\lambda} z^\alpha 
    = \frac1{\hat{z}^2}\left(\frac{z}{\hat{z}}\right)^\alpha.
\end{equation}
However, we still have one degree of freedom left in the fundamental units:  
We can choose the unit of length (or mass) so that $\hat{z}=1$ (or $\hat{\lambda}=1$).  Then all explicit constants disappear from the formula for $V$.
With this convention, $V(1) =1$ for all $\alpha$ and the potential forms an increasingly steep wall near $z=1$ as $\alpha\to\infty$.

The reduced Green function of the field equation inside a soft wall, evaluated at coincident points, was derived in \rr\citen{HardAndSoftWalls} as
\begin{equation}\label{IntroInterior210}
    g_{\kappa}(z) = \frac{1}{W} \left( F_\kappa(z)G_\kappa(z) - F_\kappa(z)^2 \frac{G_\kappa(0) - G'_\kappa(0)/\kappa}{F_\kappa(0) - F'_\kappa(0)/\kappa}\right).
\end{equation}
Here $\kappa > 0$ is the imaginary wave number, $z$ is the displacement along the axis perpendicular to the soft wall, and $W = W \left(F_\kappa,G_\kappa \right)$ is the Wronskian of $F_\kappa(z)$ and $G_\kappa(z)$, which are solutions of the differential equation
\begin{equation}\label{IntroODE}
    \left(-\frac{\partial^2}{\partial z^2} + V(z) + \kappa^2\right)y = 0,
\end{equation}
such that $F_\kappa(z)$ decays at positive infinity and $G_\kappa(z)$ is linearly independent of $F_\kappa(z)$.

By normalizing the boundary conditions of $F_\kappa(z)$ and $G_\kappa(z)$ to be
\begin{equation}\label{IntroBoundaryConds}G_\kappa(0) = 0, \qquad G'_\kappa(0) = 1, \qquad F_\kappa(0) = 1,
\end{equation}
Eq.~\eqref{IntroInterior210} was simplified\cite{swout} to 
\begin{eqnarray}\label{IntroSimplifiedInterior210}
    g_{\kappa}(z) &=& F_\kappa(z)G_\kappa(z) + \gamma(\kappa) F_\kappa(z)^2, \nonumber\\
    \gamma(\kappa) &=& \frac{1}{\kappa - F'_\kappa(0)}.
\end{eqnarray}

As shown in \rr\citen{interior}, Sec.\ II, the vacuum expectation value $\left< \phi^2 \right>$ is given by the formula $I \left[ g_{\kappa}(z) \right]$, where
\begin{equation}
    I \left[h(\kappa) \right] \equiv \frac{1}{2 \pi^2} \int _0 ^\infty d\kappa \kappa^2 h(\kappa) \frac{\sin \kappa \delta}{\kappa \delta}
\end{equation}
and $\delta$ is a point-separation parameter (see Eq.~(2.8) of \rr\citen{interior}),
which henceforth will be taken to zero in all equations where the regularization is no longer necessary.
Expressions for the components of the corresponding renormalized stress tensor inside a soft wall 
(\rr\citen{interior}, Eq.~(5.11)) 
were also derived:
\begin{eqnarray}\label{stresstensor}
    4\pi^2 \left< T^{\mu\nu}\right>_R &= &{}-g^{\mu\nu}\frac{V^2}{8}\ln{\frac{\sqrt{V}}{\mu}}
    \\ \nonumber&&{}
    - \frac{1}{2}\left( \beta+\frac{1}{12}\right)\left(\partial^\mu\partial^\nu - g^{\mu\nu}\partial^2 \right)\left(V \ln{\frac{\sqrt{V}}{\mu}} \right)   + g^{\mu\nu}\frac{V^2}{32}\\ \nonumber &&{} +\frac{1}{96}\frac{V'^2}{V}\mathop{\mathrm{diag}}\! {(1,-1,-1,1)} - \frac{1}{24}\partial_z^2\left(\frac{V''}{V} - \frac{1}{2}\frac{V'^2}{V^2} \right)\mathop{\mathrm{diag}}\! {(-\beta,\beta,\beta,\frac{1}{4})}
  \\ \nonumber &&
  {}+ 4\pi^2\left< T^{\mu\nu}\right>[I[g-\tilde{g}]].
\end{eqnarray}
(Here $\beta$  is a conformal coupling constant, and $\mu$ the arbitrary mass scale inevitably introduced in renormalization\cite{Symmetry}.)

These expressions involve terms of the form
\begin{equation}\label{IntrogIntoGintegral}
    I \left[\kappa ^ {a} f(z) \frac{\partial^b  \left(g_{\kappa}(z) -  \tilde{g}_{\kappa}(z)\right) }{\partial z^b} \right], 
\end{equation}
where $a,b \in \{0,2\}$ and the function $f(z)$ is either a constant or the soft wall \eqref{IntroSoftWallPotential}. The second-order WKB approximation of the Green function, $\tilde{g}_{\kappa}(z)$, is given in \rr\citen{interior}, Eq.~(3.1), as
\begin{equation}\label{IntroGTilde}
    \tilde{g}_\kappa (z) = \frac{1}{2 \sqrt{\kappa^2+V(z)}} - \frac{V''(z)}{16 \left(\kappa^2+V(z) \right)^{5/2}} %\nonumber\\    &\quad&
    + \frac{5V'^2(z)}{64 \left(\kappa^2+V(z)\right)^{7/2}}\,.
\end{equation}
As will be evident in the next sections, it is instructive to break up the integral \eqref{IntrogIntoGintegral} by employing a cutoff parameter $\Lambda$ and defining the functionals
\begin{eqnarray}\label{IntroIJOperators}
        I_\Lambda \left[h(\kappa) \right] &\equiv& \frac{1}{2 \pi^2} \int _0 ^\Lambda d\kappa\, \kappa^2 h(\kappa), \nonumber \\
        J_\Lambda \left[h(\kappa) \right] &\equiv& \frac{1}{2 \pi^2} \int _\Lambda ^ \infty d\kappa\, \kappa^2 h(\kappa).
\end{eqnarray}

The components of the stress tensor inside the linear and quadratic walls were computed exactly \cite{interior} using Eq.~\eqref{IntroSimplifiedInterior210}. These correspond to the only values of $\alpha$ for which exact solutions of Eq.~\eqref{IntroODE} can be expressed in terms of standard special functions, although some properties of the solutions for general $\alpha$ have been studied by
Faierman\cite{faierman}.

In order to study the Casimir effect under the soft wall model, the renormalized stress tensor should be computed for $\alpha \gg 1$. From Eq.~\eqref{IntroSoftWallPotential}, it is clear that when $\alpha$ is large, the soft wall approximates the Dirichlet wall, for which the Casimir effect was originally derived.

Since the solutions of Eq.~\eqref{IntroODE} do not have closed-form expressions for $\alpha > 2$, we must approximate $F_\kappa(z), G_\kappa(z)$ for all $\kappa \in [0,\infty)$ in order to compute terms of the form \eqref{IntrogIntoGintegral}. This is the subject of the next three sections.

The issues that arise in renormalizing the quantum theory
of the scalar field $\phi$ interacting with the
classical scalar field $V$ have been discussed in
\rr\citen{Symmetry}.
Since the main thrust of this article is the practical
problem of calculating the approximations just described, we do
not wish to belabor the renormalization theory here.  Unlike
perfectly reflecting or conducting walls, such a problem fits
fairly well into conventional quantum field theory without great
controversy.  In our application, however, we are not interested
in scattering problems, but rather in the potential $V$ as a
simple model of a gravitational field (nontrivial metric tensor).
In the cited paper\cite{Symmetry} the situation was analyzed
using Pauli--Villars regularization, a close relative of
point-splitting.  The divergences are covariantly absorbed into
coupling constants of the potential regarded as a dynamical field
in its own right, or the square of one.  Care was taken to
preserve the covariant conservation law of the stress tensor and
to document the resulting trace anomaly.  Unavoidably there are
covariant and local terms in the renormalized equation of motion
and stress tensor with numerical coefficients and an
arbitrary length scale that are not
determined by the theory.  The situation is very similar to that
for field theories in a gravitational background, as established
by Wald\cite{wald1,wald2}.  Besides Wald's papers, the work on
renormalization that has most influenced our program is that by
Mazzitelli et al.\cite{MNS}. 
(Along with other papers by the same research group\cite{pazmaz,SMA,FvMM},
the latter employs dimensional regularization.)
Many other papers are cited
in \rs\citen{swout,Symmetry}; for historical reasons we should
add a paper by Bordag and Lindig\cite{BL}.

\section{Perturbation Theory}\label{section:perturbationTheory} 

To approximate the Green function in the small $\kappa$ regime, we generate the perturbative expansions of the basis solutions $F_\kappa(z)$ and $G_\kappa(z)$ of Eq.~\eqref{IntroODE}. The coefficients of these power series are obtained from perturbation theory, as outlined in \rr\citen{swout}, Sec.\ III. There, the coefficients $F_0(z)$, $F_1(z)$, $G_0(z)$, and $G_1(z)$, in the sense of Eq.~\eqref{MethodsPerturbativeExpansions}, were computed to form first-order approximations. In particular, $F_0$ and $G_0$ are given there in terms of certain 
Bessel functions, called $k(z)$ and $i(z)$. Namely, 
\begin{equation}
 F_0(z) = {2b^b\over \Gamma(b)}\,k(z), \quad
G_0(z) = {\Gamma(b) \over 3b^b}\, i(z), 
\end{equation}
\begin{equation}
b= \frac1{\alpha+2}\,,\quad
k(z)=\sqrt{z}K_b\bigl(2bz^{1/2b}\bigr), \quad
i(z)=\sqrt{z}I_b\bigl(2bz^{1/2b}\bigr). 
\end{equation}

For the perturbative approximations to be usable as a part of a sufficiently accurate piecewise approximation, they must be carried to a rather high order. This is done by calculating the coefficients (\rr\citen{swout}, Eq.~(3.12)) using the generalization of the recursion relation (\rr\citen{swout}, Eq.~(3.8))
\begin{eqnarray}\label{MethodsPerturbationCoefficients}
    F_n(z) &=& \frac{1}{W(i,k)} \left ( k(z) \int _0 ^ z i(a)F_{n-1}(a)\, da \right. %\nonumber\\ &\quad& 
    + \left. i(z) \int _z ^ \infty k(a)F_{n-1}(a) \,da \right), \nonumber\\\nonumber\\
    G_n(z) &= &\frac{1}{W(i,k)} \left ( k(z) \int _0 ^ z i(a)G_{n-1}(a)\, da \right. %\nonumber\\ &\quad& 
    - \left.  i(z) \int _0 ^ z k(a)G_{n-1}(a)\, da \right),
\end{eqnarray}
where $W(i,k)$ is the Wronskian of $i(z)$ and $k(z)$. The coefficients $\{F_n(z)\}_{n = 2} ^ {30}$ were evaluated numerically, where the upper limit of the improper integral in Eq.~\eqref{MethodsPerturbationCoefficients} was replaced by $z_\infty = 3.4$ because {\sl Mathematica\/} cannot evaluate the integrand for large values of $a$ properly (despite its being numerically negligible). The coefficients $\{G_n(z)\}_{n = 2} ^ {30}$ were evaluated by expanding the appropriate integrands in Eq.~\eqref{MethodsPerturbationCoefficients} as Maclaurin series of order 40. Thus we obtained the 30\textsuperscript{th}-order perturbation approximations
\begin{equation}\label{MethodsPerturbativeExpansions}
    F_{\kappa \ll 1}(z) = \sum _{n = 0} ^ {30} F_n(z) \kappa ^ {2n}, \qquad% \nonumber\\
    G_{\kappa \ll 1}(z) = \sum _{n = 0} ^ {30} G_n(z) \kappa ^ {2n}.
\end{equation}

The Green function is obtained by plugging these approximations into Eq.~\eqref{IntroSimplifiedInterior210}:
 \begin{eqnarray}\label{MethodsPerturbativeGreen}
    g_{\kappa \ll 1}(z) &=& F_{\kappa \ll 1}(z) G_{\kappa \ll 1}(z) + \gamma_{\kappa \ll 1}(\kappa) F_{\kappa \ll 1}(z)^2, \nonumber\\
    \gamma_{\kappa \ll 1}(\kappa) &=& \frac{1}{\kappa - F'_{\kappa \ll 1}(0)}.
\end{eqnarray}
Next, we differentiate Eq.~\eqref{IntroSimplifiedInterior210} twice and obtain
\begin{eqnarray}\label{MethodsPerturbativeD2}
    \partial_z ^2 g_\kappa(z) &=& G_\kappa(z) F''_\kappa(z)+2 F'_\kappa(z) G'_\kappa(z) + F_\kappa(z) G''_\kappa(z) \nonumber\\ &\quad& + 2 \gamma (\kappa ) \left(F_\kappa(z) F''_\kappa(z)+F'_\kappa(z)^2\right).
\end{eqnarray}
Since $F_\kappa(z)$ and $G_\kappa(z)$ are solutions of Eq.~\eqref{IntroODE}, their second derivative is known, and we obtain
\begin{eqnarray}\label{MethodsSimplifiedPerturbativeD2}
    \frac{\left(\partial_z ^2 g_\kappa(z) \right)_{\kappa \ll 1}}{2} &=& F'_{\kappa \ll 1}(z) G'_{\kappa \ll 1}(z) + \gamma _{\kappa \ll 1}(\kappa ) \nonumber\\ &\quad& \times \left(F'_{\kappa \ll 1}(z)^2+F_{\kappa \ll 1}(z)^2 \left(\kappa ^2+V(z)\right)\right) \nonumber\\
    &\quad& + F_{\kappa \ll 1}(z) G_{\kappa \ll 1}(z)
   \left(\kappa ^2+V(z)\right),
\end{eqnarray}
in terms of $F_{\kappa \ll 1}(z), G_{\kappa \ll 1}(z)$, and their first derivatives. We find this approach to yield a more stable perturbation series than direct differentiation of the approximation \eqref{MethodsPerturbativeGreen}.

\section{The WKB Approximation}\label{section:wkbApproximation}

For large values of either $\kappa$ or $z$, the sought-after solutions are well-approximated by asymptotic expansions. The WKB approximations of simple decaying and growing solutions of Eq.~\eqref{IntroODE}, as well as the dominant part of the Green function \eqref{IntroInterior210}, were recorded (\rr\citen{interior}, Eqs.\ (A3) and (A5))  as    
\begin{eqnarray}\label{MethodsWKBOdeSolutions}
    y(z) &\sim& \frac{e^{\pm\int_{}^{z}dt[q_0(t) + q_2(t) + q_4(t) + \ldots]}}{\sqrt{q_0(z) + q_2(z) + q_4(z) + \ldots}}, \nonumber\\
    \frac{F_\kappa(z)G_\kappa(z)}{W} &\sim& \frac{1}{2}\frac{1}{q_0(z) + q_2(z) + q_4(z) + \ldots}.
\end{eqnarray}
We require that our basis solutions have the asymptotic
approximations
\begin{eqnarray}\label{MethodsWKBApproximations}
    \widehat F_{\kappa \gg 1}(z) &\equiv& \frac{e^{-\int_{}^{z}dt[q_0(t) + q_2(t) + q_4(t) + \ldots]}}{\sqrt{q_0(z) + q_2(z) + q_4(z) + \ldots}}, \nonumber\\
    \widehat G_{\kappa \gg 1}(z) &\equiv& \frac{e^{+\int_{}^{z}dt[q_0(t) + q_2(t) + q_4(t) + \ldots]}}{2\sqrt{q_0(z) + q_2(z) + q_4(z) + \ldots}},
\end{eqnarray}
with the same lower limit in both integrals. Then from Eq.~\eqref{MethodsWKBOdeSolutions} one sees that the Wronskian of the exact solutions is unity.
These solutions may not have the normalization \eqref{IntroBoundaryConds}, but since Eq.~\eqref{IntroInterior210} holds for any decaying and growing basis functions, plugging in $\widehat F_{\kappa \gg 1}(z)$ and $\widehat G_{\kappa \gg 1}(z)$ yields
\begin{eqnarray}\label{MethodsWKBGreen}
    g_{\kappa\gg1}(z) &\equiv& \frac{1}{2\left (q_0(z) + q_2(z) + q_4(z) + \ldots \right)} \nonumber\\ &\quad&+ \frac{\widehat \gamma (\kappa)e^{-2\int_{}^{z}dt[q_0(t) + q_2(t) + q_4(t) + \ldots]}}{q_0(z) + q_2(z) + q_4(z) + \ldots} \sim g_\kappa(z), \nonumber\\\nonumber\\
    \widehat \gamma (\kappa) &=& -\frac{\widehat G_{\kappa \gg 1}(0) - \widehat G_{\kappa \gg 1}'(0)/\kappa}{\widehat F_{\kappa \gg 1}(0) - \widehat F_{\kappa \gg 1}'(0)/\kappa}.
\end{eqnarray}

 As shown in \rr\citen{campbell}, the local functionals $q_n$ may be obtained from a recurrence relation. (In practice, we used a slightly different form of the recursion, found in \rr\citen{gustavus}.) The recurrence relation can be solved in a bottom-up manner using dynamic programming. With this method, the local functionals $q_1,\dots,q_n$ can be computed in $O\left(n^2\right)$ time. 

 The first five terms were kept in the prefactor and the exponential, yielding a ninth-order Fr\"{o}man approximation (\rr\citen{interior}, Appendix~A). The term $\widehat \gamma (\kappa)$ was then computed by plugging in the ninth-order Fr\"{o}man approximations of $\widehat F_{\kappa \gg 1}(z)$ and $\widehat G_{\kappa \gg 1}(z)$ and taking the limit as $z \to 0$. The second derivative of the Green function is approximated as
\begin{equation}\label{MethodsWKBD2}
    \left( \partial_z ^2 g_\kappa(z) \right)_{\kappa \gg 1} \equiv \partial_z ^2 \left( g_{\kappa\gg1}(z) \right).
\end{equation}

\section{The Approximated Green Function}\label{section:theIntermediateRegime}

 Equipped with the approximations of the last two sections, we wish to approximate the Green function for values of $\kappa$ that are too large for the perturbative expansion to be usable, but are not large enough for the WKB approximation to be valid. This process naturally contains arbitrary elements which are rooted in trial and error. These ambiguities are unavoidable and are ultimately justifiable by the results they yield (cf.\ \ref{appendix:methods}).
 %Appendix~B). 

 The Green function and its second derivative are approximated by interpolating splines in the intermediate $\kappa$ regime. For a fixed $z = z_0$, the splines are constructed to match the perturbative and WKB approximations at the boundaries of the intermediate regimes. The boundaries are defined by the endpoints $\kappa_{L_0}$, $\kappa_{R_0}$, $\kappa_{L_2}$, and  $\kappa_{R_2}$ for $ g_\kappa(z)$ and $\partial_z ^2 g_\kappa(z)$, respectively.
 
 The general form of the splines was chosen to be a simple asymptotic expansion 
\begin{eqnarray}
    g_{\kappa \approx 1}(z_0) &=& \sum _{i = 1} ^ n \frac{A_i}{\kappa^i},\label{MethodsSplineFormG} \\
    \left( \partial_z ^2 g_\kappa(z_0) \right)_{\kappa \approx 1} &=& \sum _{i = 1} ^ n \frac{B_i}{\kappa^i}.\label{MethodsSplineFormD2G}
\end{eqnarray}

\noindent The coefficients $\{A_i\},\{B_i\}$ were selected to satisfy
\begin{eqnarray}\label{MethodsSplineConds}
 \partial^m_\kappa g_{\kappa \approx 1} (z_0) \bigg|_{\kappa_{L_0}} &=& \partial^m_\kappa g_{\kappa \ll 1} (z_0) \bigg|_{\kappa_{L_0}}, \nonumber\\ \partial^m_\kappa \left( \partial_z ^2 g_\kappa(z_0) \right)_{\kappa \approx 1} \bigg|_{\kappa_{L_2}} &=& \partial^m_\kappa \left( \partial_z ^2 g_\kappa(z_0) \right)_{\kappa \ll 1} \bigg|_{\kappa_{L_2}}, \nonumber\\
\partial^m_\kappa g_{\kappa \approx 1} (z_0) \bigg|_{ \kappa_{R_0}} &=& \partial^m_\kappa g_{\kappa \gg 1} (z_0) \bigg|_{\kappa_{R_0}}, \nonumber\\ \partial^m_\kappa \left( \partial_z ^2 g_\kappa(z_0) \right)_{\kappa \approx 1} \bigg|_{\kappa_{R_2}} &=& \partial^m_\kappa \left( \partial_z ^2 g_\kappa(z_0) \right)_{\kappa \gg 1} \bigg|_{ \kappa_{R_2}},
\end{eqnarray}
for all $m = 0,\dots,\frac{n-2}{2}$.

\begin{figure}[h!] % This is how you include a figure
	\centering    % This centers it
	\includegraphics[width=\linewidth]{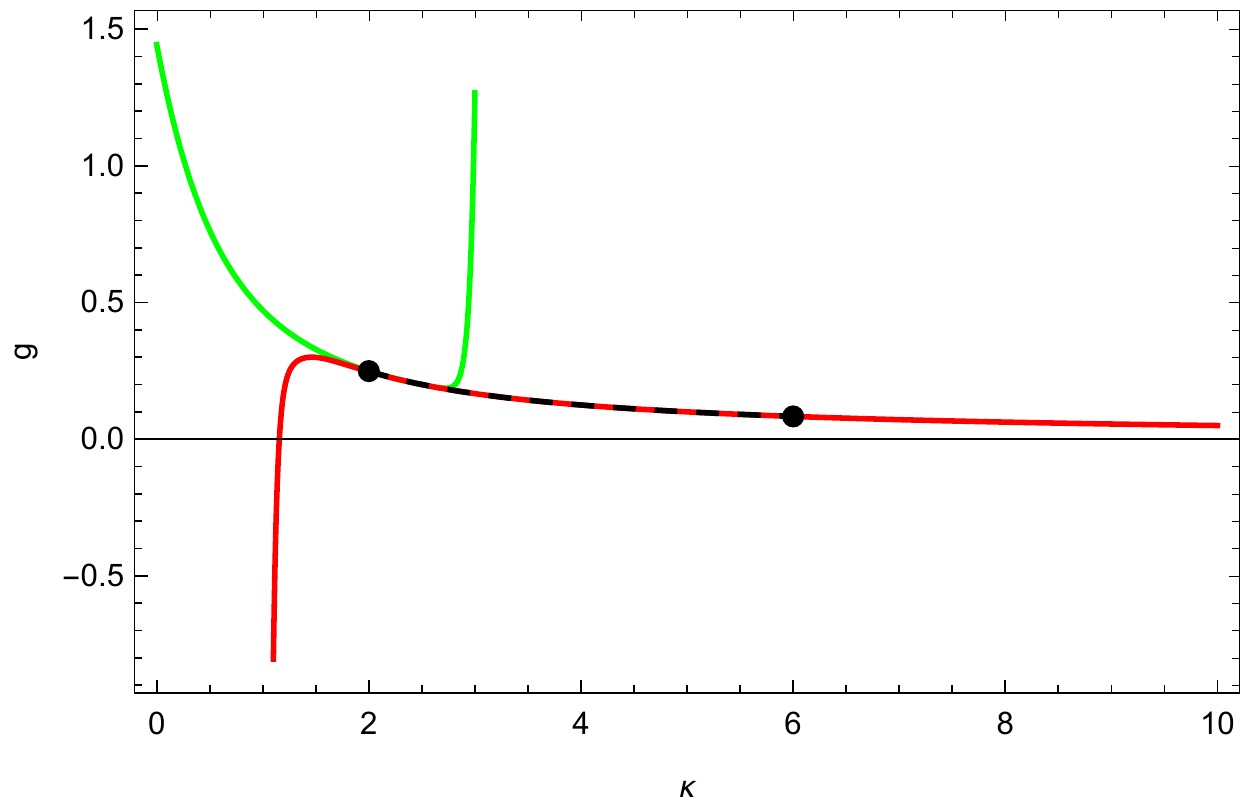}
	\caption{The perturbative expansion \eqref{MethodsPerturbativeGreen} (in green), the WKB approximation \eqref{MethodsWKBGreen} (in red), and the spline \eqref{MethodsSplineFormG} (in black dashes) for $\alpha=6$ and $z = 0.01$. The black dots mark $\kappa_{L_0} = 2$ and $\kappa_{R_0} = 6$. The perturbative expansion clearly diverges as $\kappa$ grows, while the WKB approximation is accurate for moderate values of $\kappa$. All graphics in this paper were prepared with {\sl Mathematica\/}.}
	\label{fig:gApprox}
\end{figure}

\begin{figure}[h!] % This is how you include a figure
	\centering    % This centers it
\includegraphics[width=\linewidth]{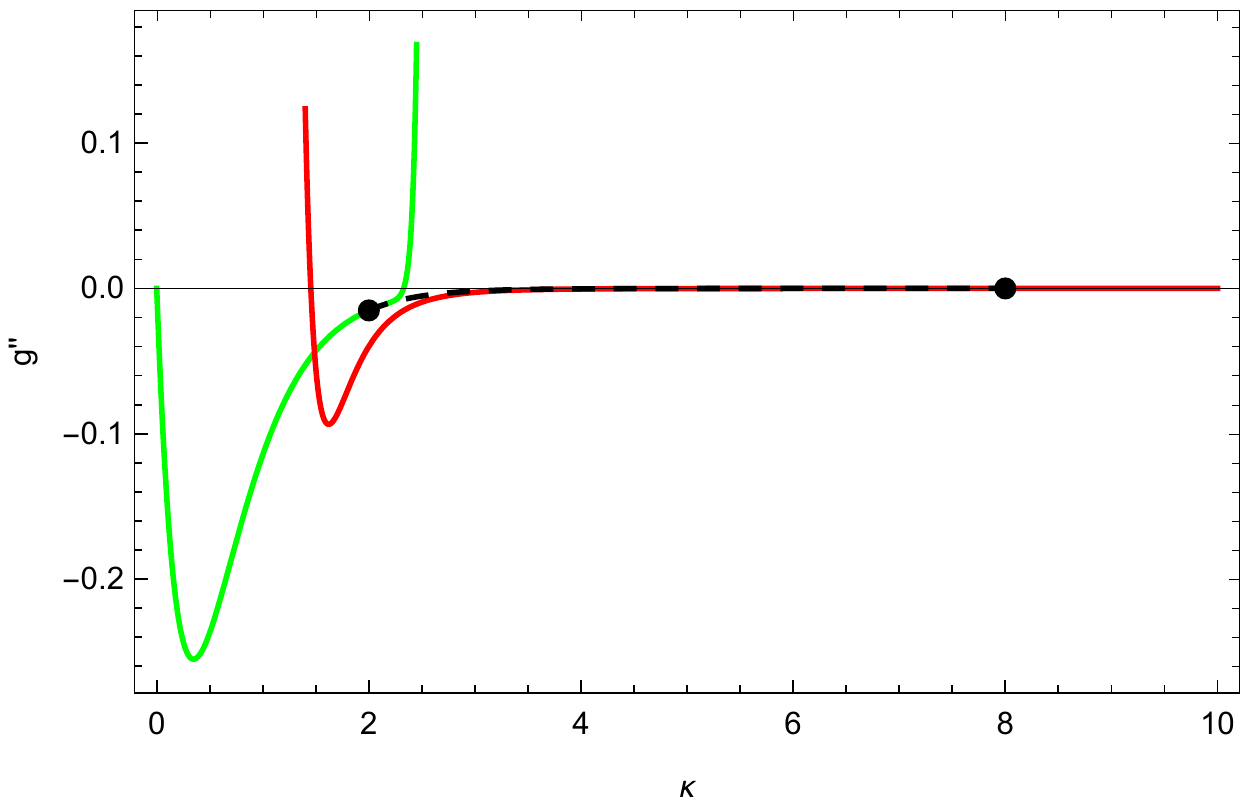}
	\caption{The perturbative expansion \eqref{MethodsSimplifiedPerturbativeD2} (in green), the WKB approximation \eqref{MethodsWKBD2} (in red), and the spline \eqref{MethodsSplineFormD2G} (in black dashes) for $\alpha=6$ and $z = 0.01$. The black dots mark $\kappa_{L_2} = 2$ and $\kappa_{R_2} = 8$.}
	\label{fig:D2gApprox}
\end{figure}

The endpoints were chosen arbitrarily to be within the domain of validity of their corresponding regimes while also being in close proximity to each other. This criterion may be formulated in a well-defined manner using the remainders of the perturbation and WKB approximations. Since these are yet to be calculated, we implement the endpoint selection by qualitative analysis of the behavior of both approximations. The spline $g_{\kappa \approx 1}(z_0)$, shown in Fig.~\ref{fig:gApprox}, was constructed using $\kappa_{L_0} = 2$ and $\kappa_{R_0} = 6$. The spline $\left( \partial_z ^2 g_\kappa(z_0) \right)_{\kappa \approx 1}$, shown in Fig.~\ref{fig:D2gApprox}, was constructed using $\kappa_{L_2} = 2$ and $\kappa_{R_2} = 8$. Both splines were taken to be of order $n = 8$.

Using the approximations \eqref{MethodsPerturbativeGreen}, \eqref{MethodsWKBGreen}, and \eqref{MethodsSplineFormG}, we approximate the Green function in all $\kappa$ regimes as

\begin{equation}\label{MethodsPiecewiseGreen}
    \overline{g}_{\kappa}(z_0) \equiv  
    \begin{cases}
    g_{\kappa \ll 1}(z_0) & \kappa < \kappa_{L_0}, \\
    g_{\kappa \approx 1}(z_0) & \kappa_{L_0} \leq \kappa \leq \kappa_{R_0}, \\
    g_{\kappa \gg 1}(z_0) & \kappa > \kappa_{R_0}.
\end{cases}
\end{equation}
Similarly, the approximations \eqref{MethodsSimplifiedPerturbativeD2}, \eqref{MethodsWKBD2}, and \eqref{MethodsSplineFormD2G} yield the piecewise approximation
\begin{equation}\label{MethodsPiecewiseD2Green}
    \partial_z ^2 \overline{g}_{\kappa}(z_0) \equiv  
    \begin{cases}
    \left( \partial_z ^2 g_\kappa(z) \right)_{\kappa \ll 1} & \kappa < \kappa_{L_2}, \\
    \left( \partial_z ^2 g_\kappa(z) \right)_{\kappa \approx 1} & \kappa_{L_2} \leq \kappa \leq \kappa_{R_2}, \\
   \left( \partial_z ^2 g_\kappa(z) \right)_{\kappa \gg 1} & \kappa > \kappa_{R_2}.
\end{cases}
\end{equation}

To verify the accuracy of the piecewise approximations, we compare them to their numerical counterparts. These are obtained by solving Eq.~\eqref{IntroODE} numerically with the boundary conditions \eqref{IntroBoundaryConds} to obtain the numerical basis functions $\underline F_\kappa(z)$ and $\underline G_\kappa(z)$. This process is outlined in \ref{appendix:numericalApprox}.

\begin{figure}[h!]
	\centering
	\includegraphics[width = \linewidth]{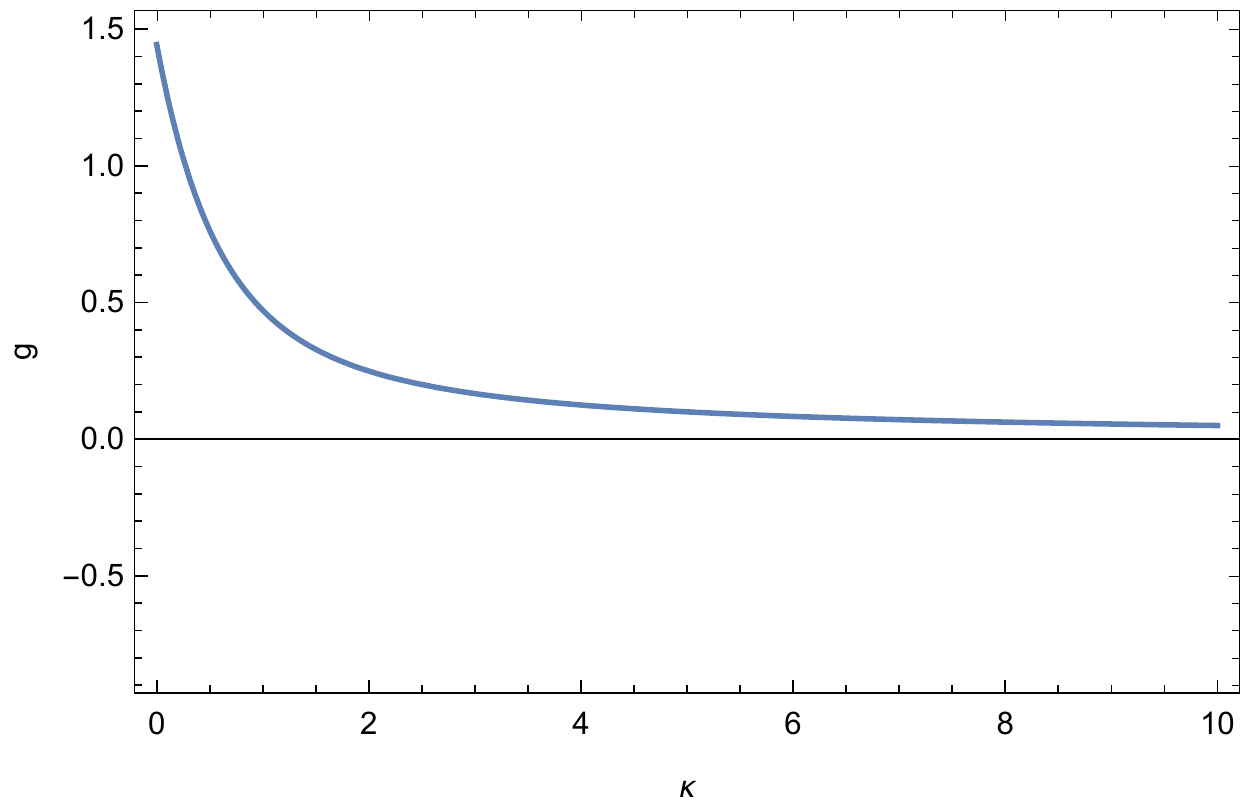}
	\caption{The numerical Green function $\underline g_\kappa(z)$ for $V(z) = z^6$ and $z = 0.01$.}
	\label{fig:CHOL_g}
\end{figure}

\begin{figure}[h!]
	\centering
	\includegraphics[width = \linewidth]{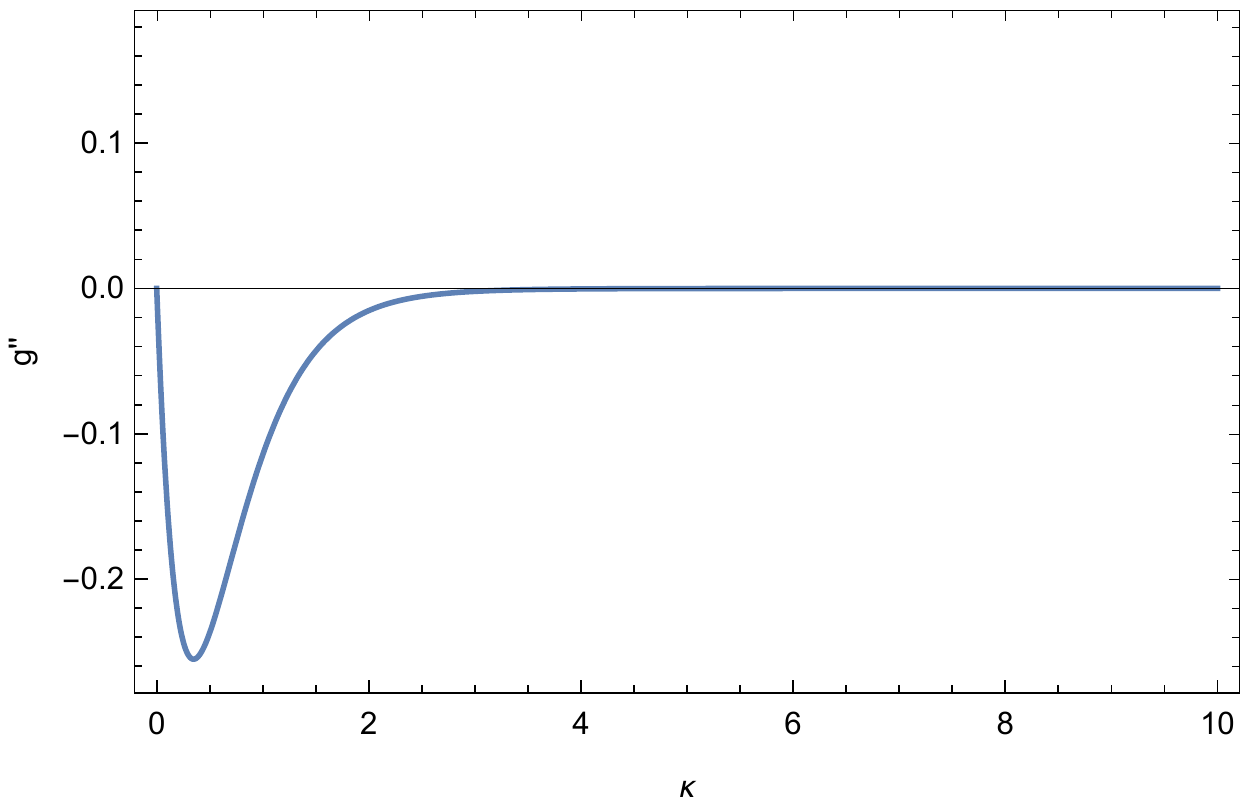}
	\caption{The second derivative of the numerical Green function $\underline {\partial_z^2 g}_\kappa(z)$ for $V(z) = z^6$ and $z = 0.01$. %To minimize numerical error, $\underline {\partial_z^2 g}_\kappa(z)$ was calculated from the basis functions and their first derivatives similarly to Eq.~\eqref{MethodsSimplifiedPerturbativeD2}.}
	}
	\label{fig:CHOL_ddg}
\end{figure}

The numerical Green function $\underline g_\kappa(z)$ and its second derivative $\underline {\partial_z^2 g}_\kappa(z)$ are plotted in Figs.~\ref{fig:CHOL_g} and \ref{fig:CHOL_ddg}. They were computed from the numerical basis solutions $\underline F_\kappa(z)$ and $\underline G_\kappa(z)$ using Eq.~\eqref{IntroSimplifiedInterior210} and the numerical equivalent of Eq.~\eqref{MethodsSimplifiedPerturbativeD2}. The derivatives $\underline F'_\kappa(z)$ and $\underline G'_\kappa(z)$ were approximated by sixth-order finite difference formulae \cite{finiteDifferenceCoeffs, finiteDifferenceAlgorithm}.

The piecewise and numerical approximations presented here will be used in the next section to compute the stress tensor inside a soft wall. The terms of the form \eqref{IntrogIntoGintegral} are approximated analytically using the functionals \eqref{IntroIJOperators} as
%\begin{eqnarray}\label{MethodsApproxIop}
%   \overline{I}\left[\partial_z^b \left[ g - \tilde{g} \right] \right] &\equiv& I_{10}\left[\overline{\partial_z^b  g}_\kappa(z) \right] -  I_{10}\left[\partial_z^b \tilde{g}_\kappa(z) \right] \nonumber\\ &\quad& + J_{10}\left[\left(\partial_z^b g_\kappa(z) \right)_{\kappa \gg 1} - \partial_z^b  \tilde{g}_\kappa(z) \right],
%\end{eqnarray}
\begin{eqnarray}\label{MethodsApproxIop}
   \overline{I}\left[\partial_z^b \left[ g - \tilde{g} \right] \right] &\equiv& J_{10}\left[\left(\partial_z^b g_\kappa(z) \right)_{\kappa \gg 1} - \partial_z^b  \tilde{g}_\kappa(z) \right] \nonumber\\ &\quad& + I_{10}\left[\overline{\partial_z^b  g}_\kappa(z) \right] -  I_{10}\left[\partial_z^b \tilde{g}_\kappa(z) \right],
\end{eqnarray}
and computed numerically as 
%\begin{eqnarray}\label{MethodsNumIop}
%   \underline{I}\left[\partial_z^b \left[ g - \tilde{g} \right] \right] &\equiv& I_{25}\left[\underline{\partial_z^b g}_\kappa(z) \right] -  I_{25}\left[\partial_z^b \tilde{g}_\kappa(z) \right] \nonumber \\ &\quad& + J_{25}\left[\left(\partial_z^b g_\kappa(z) \right)_{\kappa \gg 1} - \partial_z^b  \tilde{g}_\kappa(z) \right].
%\end{eqnarray}
\begin{eqnarray}\label{MethodsNumIop}
   \underline{I}\left[\partial_z^b \left[ g - \tilde{g} \right] \right] &\equiv& J_{25}\left[\left(\partial_z^b g_\kappa(z) \right)_{\kappa \gg 1} - \partial_z^b  \tilde{g}_\kappa(z) \right] \nonumber \\ &\quad& + I_{25}\left[\underline{\partial_z^b g}_\kappa(z) \right] -  I_{25}\left[\partial_z^b \tilde{g}_\kappa(z) \right].
\end{eqnarray}

Because of the defined behavior of the piecewise approximations \eqref{MethodsPiecewiseGreen} and \eqref{MethodsPiecewiseD2Green} in the large $\kappa$ region, $\overline{I}$ is insensitive to the choice of the cutoff parameter, as long as $\overline{\Lambda} > \kappa_{R_0}, \kappa_{R_2}$. Therefore, we arbitrarily take $\overline{\Lambda} = 10$.

%The numerical cutoff parameter $\underline{\Lambda} = 25$ reflects the region where the numerical solution is well-behaved. As $\kappa$ grows beyond $\kappa = 25$, the numerical second derivative starts decreasing unexpectedly due to increasing error.

The numerical cutoff parameter $\underline{\Lambda} = 25$ reflects the region where the numerical solution is valid. For values slightly larger than $\kappa = 25$, the second derivative of the computed Green function starts decreasing. Although minuscule, this behavior is undesirable and disagrees with the expected qualitative behavior of $\partial_z^2 g_\kappa(z)$.

\section{The Energy Density Inside the Sextic Wall}\label{section:theEnergyDensityInside}

 Using the scheme developed in the last section and the stress tensor formula \eqref{stresstensor}, we approximate the renormalized energy density inside the sextic wall, $V(z) = z^6$. This exponent is large enough to exhibit reasonably wall-like 
behavior, and at much larger values one begins to encounter numerical difficulties. The arbitrary mass scale $\mu$ is taken to be 1.

Following the notation introduced in \rr\citen{swout}, we define $\beta \equiv \xi -\frac{1}{4}$, where $\xi$ is the conformal parameter. Plugging the sextic potential into %\cite[Eq.~(5.11)]{interior} 
\eqref{stresstensor}
yields
\begin{eqnarray}\label{ResultsT00R}
    \left<T^{00} \right>_R &=& \left<T^{00} \right> \left[ I\left[g - \tilde{g} \right]\right] + \frac{96 \beta -z^8 \left(528 \beta +z^8+32\right)}{128 \pi ^2
   z^4} \nonumber \\ &\quad& +\frac{3 z^4 \left(-120 \beta +z^8-10\right) \ln z}{32 \pi ^2},
\end{eqnarray}
 where the first term is given by  Eq.~(6.3) of \rr\citen{interior} as
\begin{equation}\label{ResultsT00Iop}
    \left<T^{00} \right> \left[ I\left[g - \tilde{g} \right]\right] =  -\left( I\left[\frac{\kappa^2}{3} \left[ g - \tilde{g} \right] \right] +  I\left[\beta \partial_z^2 \left[ g - \tilde{g} \right] \right] \right).
\end{equation}
 We compute this term using the analytical and numerical functionals $\overline{I}$ and $\underline{I}$ defined in Eqs.~\eqref{MethodsApproxIop} and \eqref{MethodsNumIop}, respectively. The corresponding integrals were evaluated numerically for 20 evenly-spaced values of $z$ and interpolated. 

\begin{figure}[h!]
	\centering
	\includegraphics[width=\linewidth]{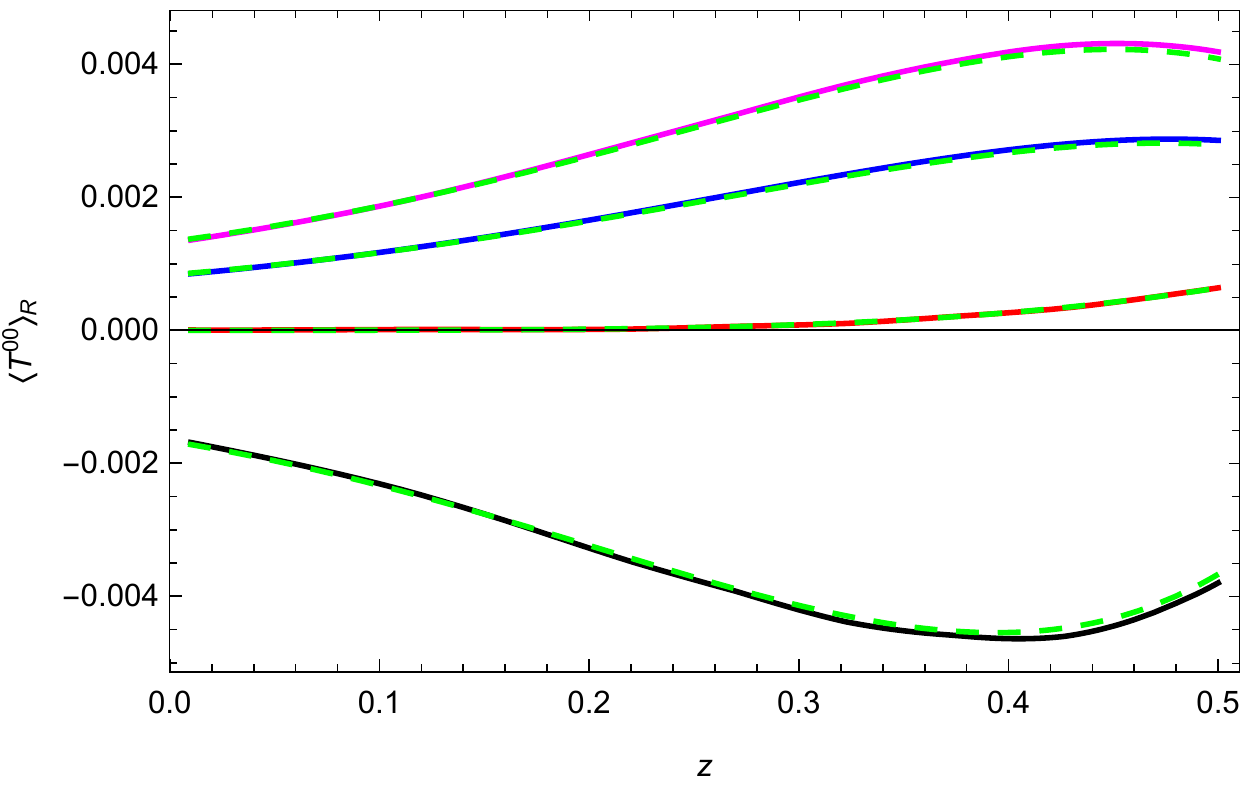}
    \caption{The approximated energy density inside the sextic wall for $\beta = 1/20$, $\beta = 0$, $\beta = -1/12$ (the conformal value), and $\beta = -1/4$ (the minimal coupling value), from top to bottom. The dashed green curves are the corresponding numerical energy densities.}
    \label{fig:T00Small}
\end{figure}

\begin{figure}[h!]
	\centering
	\includegraphics[width=\linewidth]{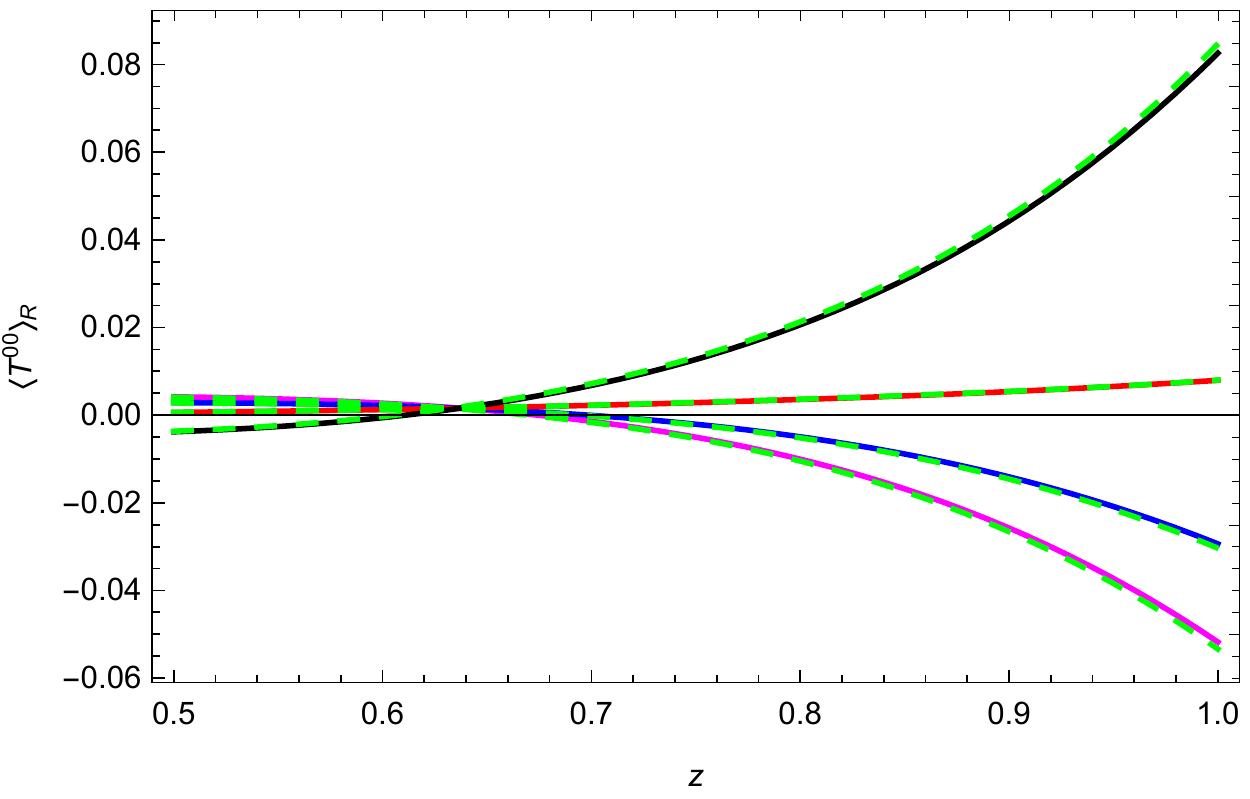}
    \caption{The energy density inside the sextic wall for large $z$. The energy density clearly changes signs in the large $z$ region.}
    \label{fig:T00Large}
\end{figure}

\begin{figure}[h!]
	\centering
    \includegraphics[width=\linewidth]{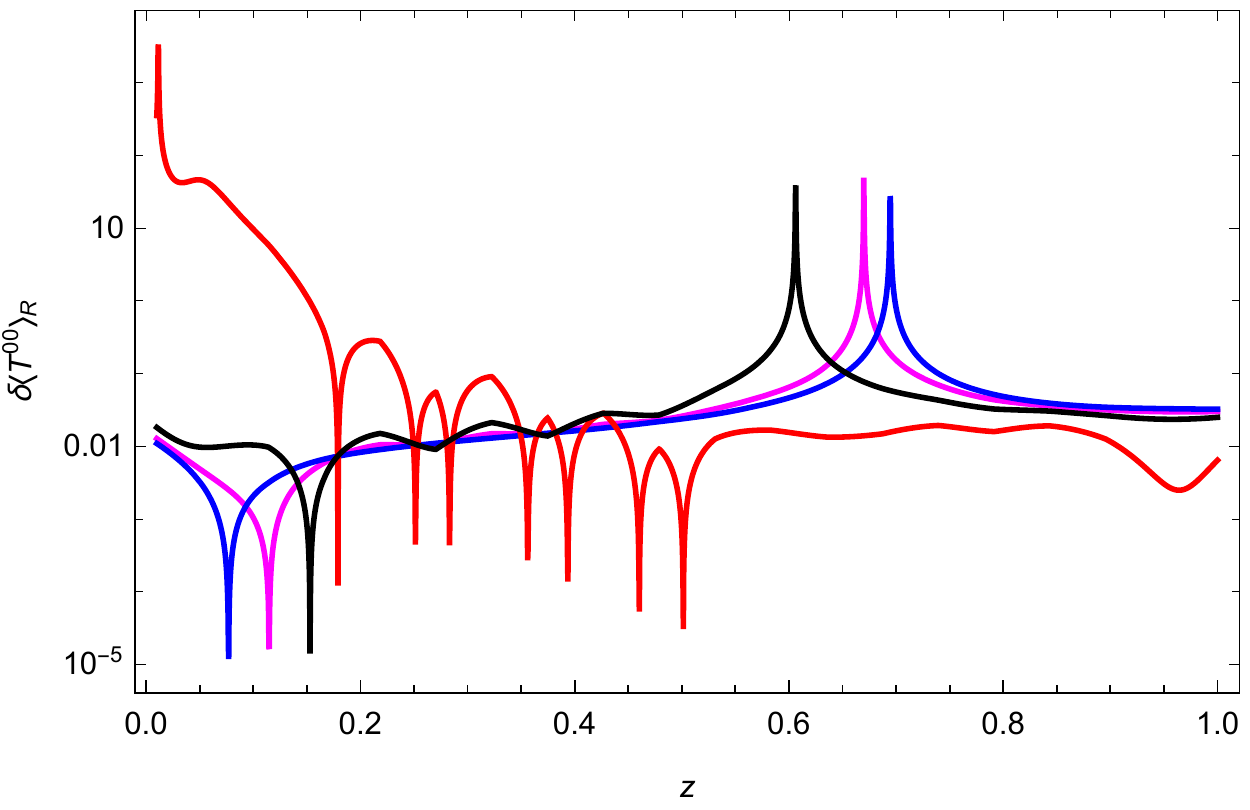}
    \caption{The relative errors in the approximated energy density inside the sextic wall. The curves show $\beta = 1/20$ in magenta, $\beta = 0$ in blue, $\beta = -1/12$ (the conformal value) in red, and $\beta = -1/4$ (the minimal coupling value) in black. The peaks occur where the numerical energy densities vanish; the absolute error is small there.}
    \label{fig:T00RelErrors}
\end{figure}

Figs.~\ref{fig:T00Small} and \ref{fig:T00Large} show the energy density for various conformal parameters in the small and large $z$ regimes. The corresponding relative errors are shown in Fig.~\ref{fig:T00RelErrors}.

\section{The Pressure Inside the Sextic Wall}\label{section:thePressureInside}

Similarly to the energy density, plugging the sextic wall $V(z) = z^6$ and the arbitrary mass scale $\mu = 1$ into \eqref{stresstensor} yields
\begin{equation}\label{ResultsTzzR}
\left<T^{zz} \right>_R = \left<T^{zz} \right> \left[ I\left[g - \tilde{g} \right]\right] %\nonumber \\ &\quad& 
+ \frac{z^{16}+12 z^8-24}{128 \pi ^2 z^4}-\frac{3 z^{12} \ln z}{32 \pi ^2},
\end{equation}
where the first term is given by Eq.~(2.13d) of \rr\citen{interior} as
\begin{equation}\label{ResultsTzzIop}
\left<T^{zz} \right> \left[ I\left[g - \tilde{g} \right]\right] =   I\left[\frac{1}{4} \partial_z^2 \left[ g - \tilde{g} \right] \right] - I\left[\left( \kappa^2 + z^6 \right) \left[ g - \tilde{g} \right] \right],
\end{equation}
 and is again computed using $\overline{I}$ and $\underline{I}$.

\begin{figure}[h!]
	\centering
    \includegraphics[width=\linewidth]{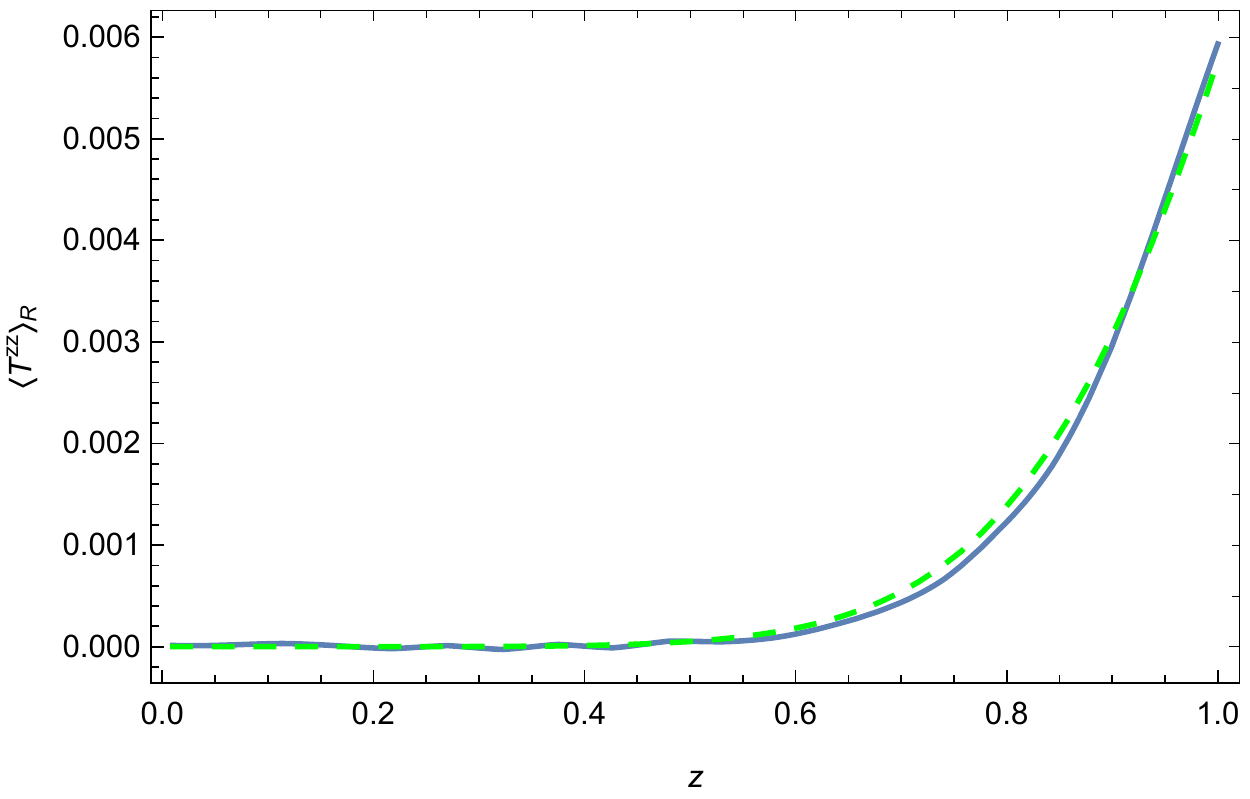}
    \caption{The approximated pressure (in light blue) and its numerical counterpart (in green dashes) inside the sextic wall. %The approximated pressure for small $z$ is dominated by the error introduced by the splines in the approximations \eqref{MethodsPiecewiseGreen} and \eqref{MethodsPiecewiseD2Green}.}
    }
    \label{fig:Tzz}
\end{figure}

\begin{figure}[h!]
	\centering
    \includegraphics[width=\linewidth]{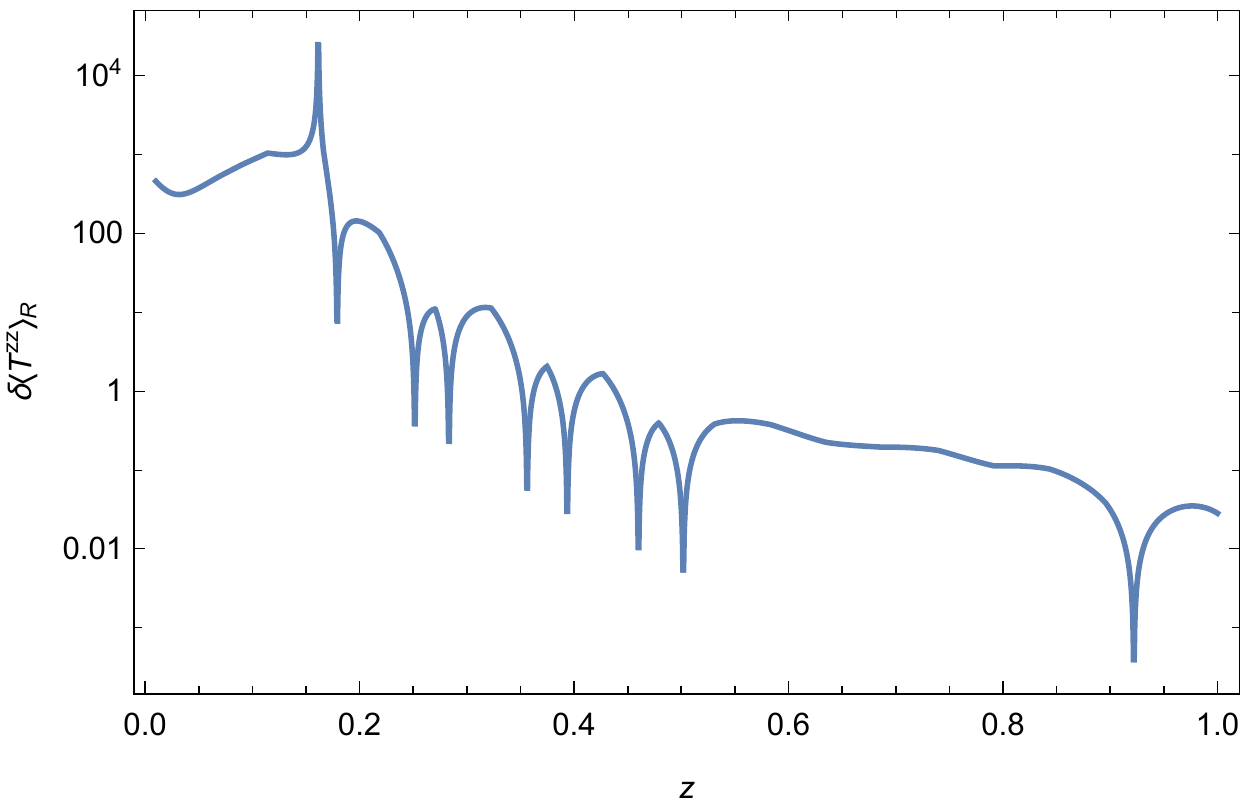}
    \caption{The relative error in the approximated pressure inside the sextic wall.}
    \label{fig:TzzRelErrors}
\end{figure}

As evident in Figs.~\ref{fig:Tzz} and \ref{fig:TzzRelErrors}, the approximated pressure is inaccurate in the small $z$ regime, despite being numerically small. Had the range of Fig.~\ref{fig:Tzz} been restricted to that of Fig.~\ref{fig:T00Small}, the curves would seem drastically different. In this region, the pressure is significantly smaller than the typical error in the approximations \eqref{MethodsPiecewiseGreen} and \eqref{MethodsPiecewiseD2Green}, rendering its accurate computation unfeasible.
 
\section{Conservation Laws}\label{section:conservationLaws}

Since the trace identity and conservation law associated with $\left< T ^ {\mu \nu}\right>_R$ depend on the vacuum expectation value of $\phi^2$, its renormalization must be consistent with the renormalization of the stress tensor %\cite[Eq.~(5.11)]{interior}. 
\eqref{stresstensor}.
We adapt the general prescription given in \rr\citen{Symmetry} to the model \eqref{MethodsGeneralizedKGLagrangian}. This Lagrangian density was obtained from Eq.~(5) of \rr\citen{Symmetry} by taking $\lambda = 1$. Following our convention, we also fix the arbitrary mass scale $\mu = 1$. With the $C_j = 0$, the renormalized field squared is given above Eq.~(36) of \rr\citen{Symmetry} as
$$\left< \phi^2 \right>_R = \frac{1}{16 \pi^2} \left(V \ln V - V + O\left(V^{-1}\right)\right).$$

As discussed on p.~10 of \rr\citen{Symmetry}, this renormalization process is not unique. Addition of $\Upsilon V$, for any constant $\Upsilon$, results in a conserved stress tensor. Therefore, we take $\Upsilon = 1$ to cancel the term proportional to $V$. Comparing this to Eq.~(5.2) of \rr\citen{interior} (with $\delta = 0$), we identify the $O\left( V^{-1} \right)$ term with $I \left[g_\kappa(z) - g_\kappa^{(0)}(z) \right]$, where $$g_\kappa^{(0)}(z) = \frac{1}{2 \sqrt{\kappa ^2 + V(z)}}$$ is the zeroth-order WKB approximation of $g_\kappa(z)$ (\rr\citen{interior}, Appendix~A).
 We obtain the renormalization
\begin{equation}\label{MethodsRenormalizedPhiSquared}
    \left< \phi^2 \right>_R = I \left[g_\kappa(z) - g_\kappa^{(0)}(z) \right] + \frac{V \ln V}{16 \pi^2}.
\end{equation}
For the sextic wall $V(z) = z^6$, we obtain
\begin{equation}\label{ResultsRenormalizedPhiSquared}
     \left< \phi^2 \right>_R \equiv I_R \left[ g_\kappa(z) \right] = I \left[g_\kappa(z) - \frac{1}{2 \sqrt{\kappa^2 + z^6}}\right] + \frac{3 z^6 \ln z}{8 \pi ^2}.
\end{equation}

\begin{figure}[h!]
	\centering
	\includegraphics[width=\linewidth]{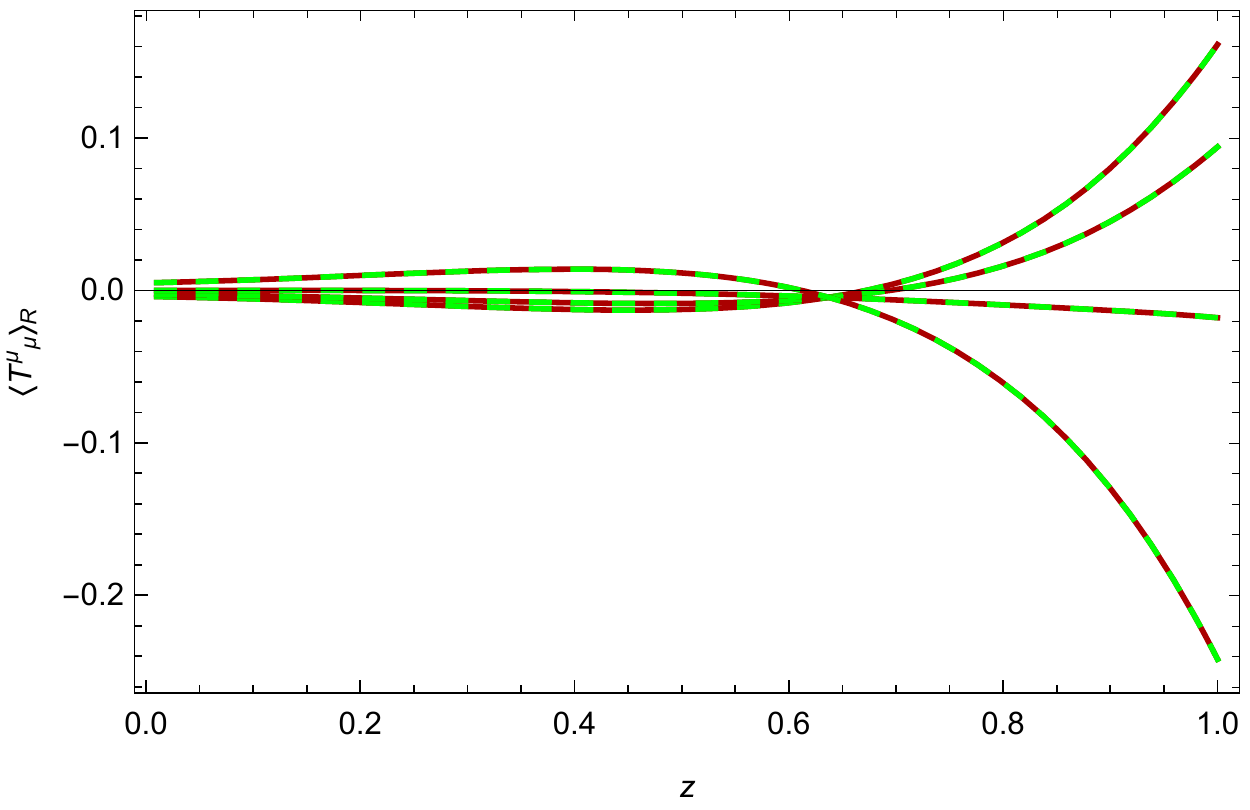}
	\caption{The approximated trace (in dark red) and the trace identity \eqref{MethodsTraceID} with $I_R \mapsto \overline{I}_R$ (in green dashes) inside the sextic wall. The curves show $\beta = 1/20, 0, -1/12, -1/4$, from top-right to bottom-right.}
   \label{fig:trace}
\end{figure}

This allows us to renormalize the trace identity and the conservation law. The trace identity (Eq.~(5.10) of \rr\citen{interior}) is renormalized as
\begin{equation}\label{MethodsTraceID}
    \left< \tensor{T}{^\mu_\mu}\right>_R + V I_R - 3 \left(\xi - \frac{1}{6} \right)\partial_z^2 I_R = \frac{1}{32 \pi^2} \left( V^2-\frac{1}{3}V'' \right).
\end{equation}
The consistency of our results with this identity is demonstrated in Fig.~\ref{fig:trace}.

\begin{figure}[h!]
	\centering
	\includegraphics[width=\linewidth]{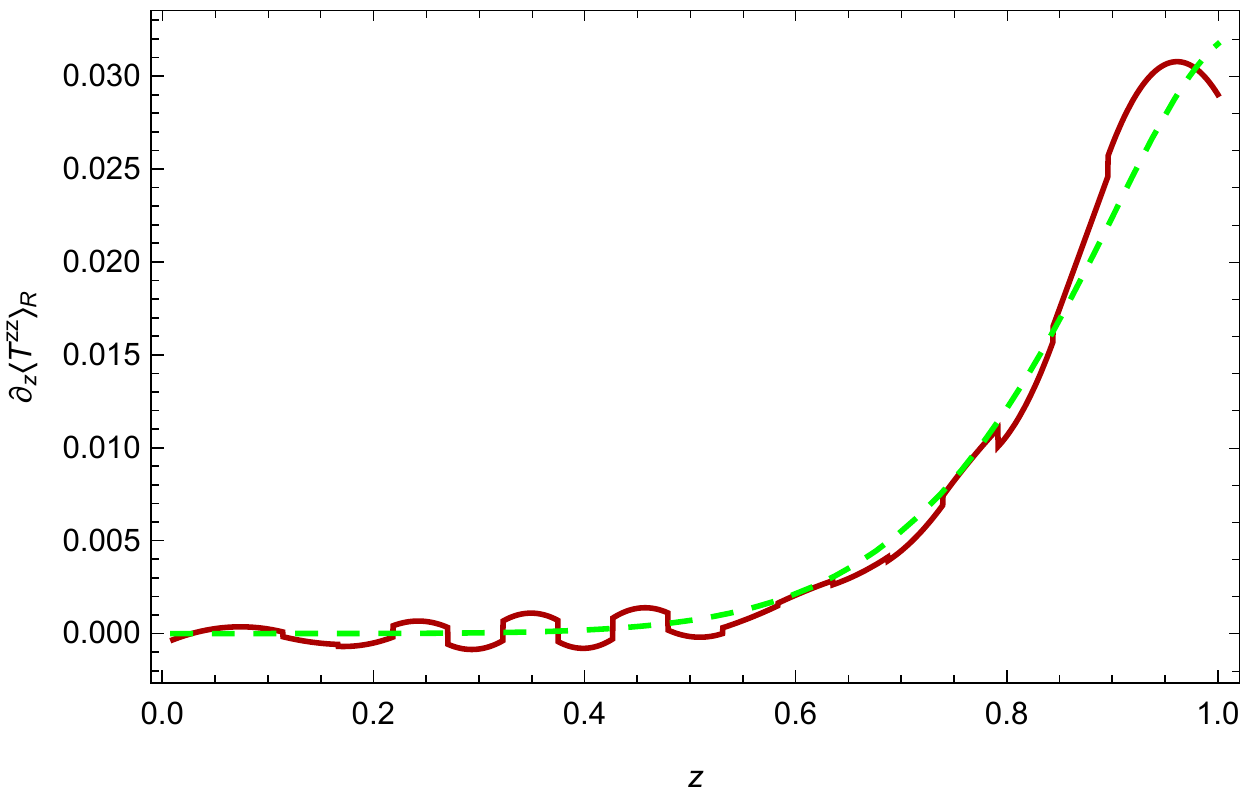}
	
	\caption{The approximated divergence (in dark red) and $-\overline{I}_R V'/2$ of Eq.~\eqref{MethodsDivergenceID} (in green dashes) inside the sextic wall. The apparent ``wiggles" are an artifact of the spline approximation.%According to the renormalized conservation law \eqref{MethodsDivergenceID}, the curves should coincide.}
	}
   \label{fig:divergence}
\end{figure}

Since $\left< T^{xx}\right>_R = \left< T^{yy}\right>_R = -\left< T^{00}\right>_R$, the divergence $\partial_{\mu} \left< T^{\mu\nu}\right>_R$ simplifies to $\partial_z \left< T^{zz}\right>_R$. Therefore, the renormalized conservation law 
\[
    \partial_\mu \left< T^{\mu\nu}\right>_R + \frac{1}{2} \partial^\nu V \left< \phi^2 \right>_R = 0,
\]
simplifies to 
\begin{equation}\label{MethodsDivergenceID}
    \partial_z \left< T^{zz}\right>_R = -\frac{V'}{2} I_R.
\end{equation}
Fig.~\ref{fig:divergence} shows both sides of the renormalized conservation law. 

\section{Reproducing the Energy Density for the Quadratic Wall}\label{section:reproducingTheQuadratic}

To further verify the approximations \eqref{MethodsPiecewiseGreen} and \eqref{MethodsPiecewiseD2Green}, we reproduce the energy density for the quadratic wall, which was computed from the exact Green function in \rr\citen{interior}. As in the sextic wall, we choose the endpoints to be in the domain of validity of the perturbative and WKB expansions. These were taken to be $\kappa_{L_0} = 1$ and $\kappa_{R_0} = 6$ for the Green function, and  $\kappa_{L_2} = 1$ and  $\kappa_{R_2} = 8$ for its second derivative. We calculate the energy density using Eqs.~\eqref{ResultsT00R} and \eqref{ResultsT00Iop}, as before.

\begin{figure}[h!]
	\centering
	\includegraphics[width=\linewidth]{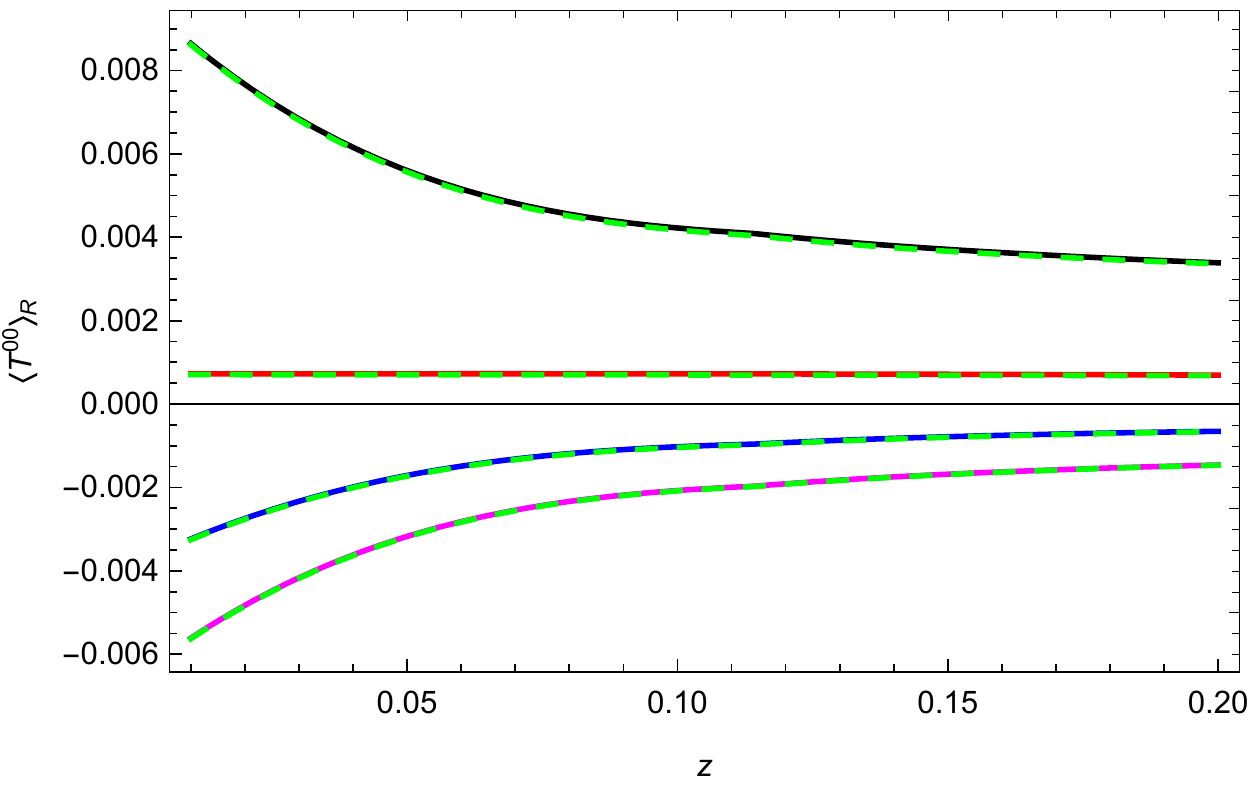}
	\caption{The approximated energy density for $\alpha = 2$ obtained from $\overline{I}$ for $\beta = 1/20, 0, -1/12, -1/4$, from bottom to top (in solid curves). The dashed green curves are the corresponding energy densities from the data of \cite[Fig.~4]{interior}.}
   \label{fig:T00quadraticwall}
\end{figure}

\begin{figure}[h!]
	\centering
	\includegraphics[width=\linewidth]{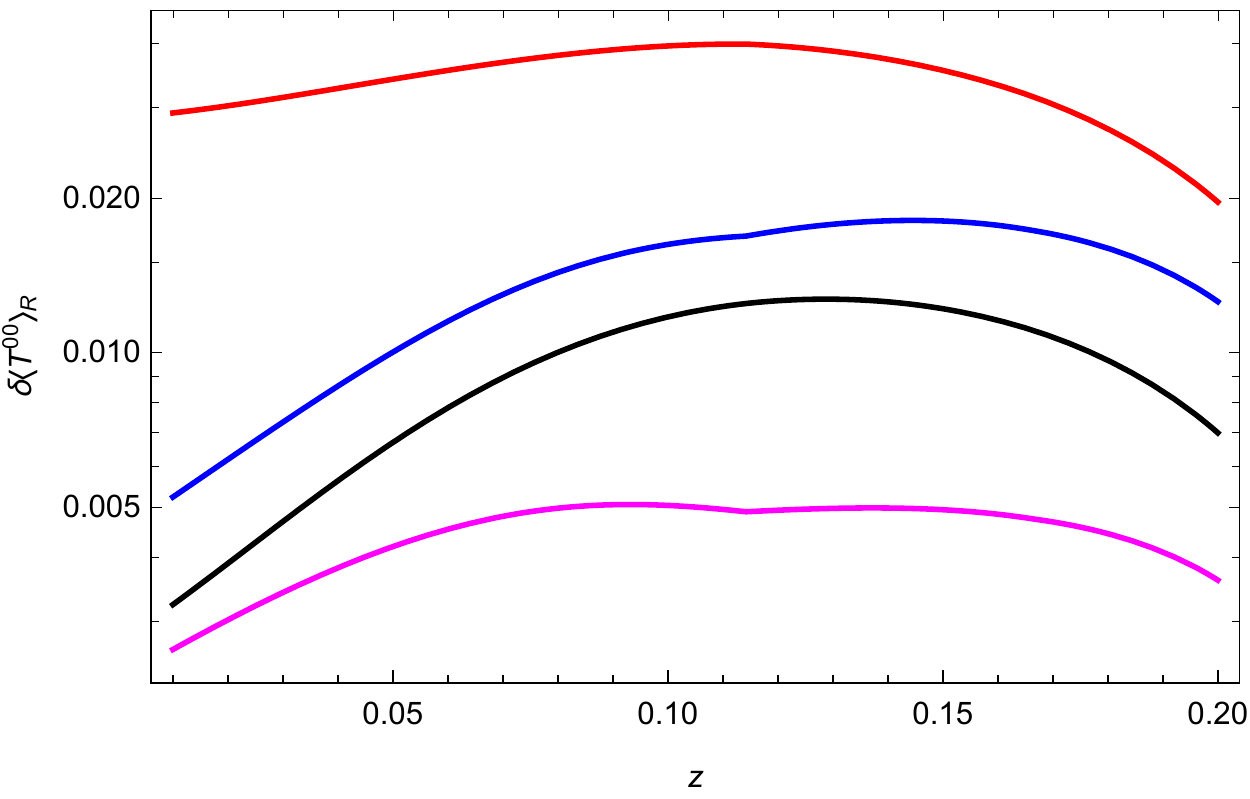}
	\caption{The relative errors in the approximated energy density inside the quadratic wall. The curves show $\beta = 1/20, -1/4, 0, -1/12$, from bottom to top (the colors are the same as in Fig.~\ref{fig:T00quadraticwall}).}
   \label{fig:T00RelErrorsquadraticwall}
\end{figure}

For comparison purposes, the data associated with Fig.~4 of \rr\citen{interior} were obtained from the authors. Figs.~\ref{fig:T00quadraticwall} and \ref{fig:T00RelErrorsquadraticwall} clearly demonstrate the consistency of our approximation with the calculation conducted in \rr\citen{interior}. Certain technical problems in the 
numerical work in \rr\citen{interior} (for reasons completely 
unrelated to any uncertainty in the present work) needed to be 
resolved by an \emph{ad hoc} fitting of one constant 
(\rr\citen{interior}, Eq.~(7.12)).
  Therefore, the observed 
consistency serves as a welcome confirmation of that calculation as well as ours.

\section{New Conclusions}\label{section:conclusion}

The methodologies developed in this paper successfully generalize the analysis and results given in \rr\citen{interior} to arbitrarily ``hard" power walls. Using high order perturbative and WKB expansions, we approximated the Green function and its relevant derivatives to high accuracy in the small and large $\kappa$ regimes.

These approximations were used to interpolate the functions in the intermediate regime. By constructing a spline that matches the perturbative and WKB expansions at appropriate endpoints, we obtained piecewise analytical approximations of the Green function and its second derivative. These approximations were then used to compute the renormalized stress tensor inside the sextic wall.

To validate the approximated results, we compared them to their numerical counterparts. The numerical approximations were obtained by discretizing the equation of motion of the scalar field $\phi$ and solving the resulting matrix equation. The solution process was optimized by utilizing techniques of numerical linear algebra and parallel computing. The approximated energy density, and to a reasonable extent the pressure, agree with their numerical counterparts, indicating their validity. The approximated stress tensor also satisfies the renormalized trace identity and conservation law. 

To further verify the validity of the developed approximation scheme, we reproduced the energy density for the quadratic wall, which was computed exactly in \rr\citen{interior}. 
Our results agree with Fig.~4 of that reference within reasonable error.

The analytical and numerical methods are complementary.  They tend to fail or become problematical in different regimes. Furthermore, both approaches require \emph{ad hoc} elements, and their agreement reassures the obtained results. The adaptivity of the analytical approach to arbitrary orders of accuracy and the insight it grants into the qualitative behavior of the Green function are advantages over the numerical calculations.

In future work, we hope to generalize and improve the approximation method in the intermediate $\kappa$ regime. The current method relies on fixing arbitrary endpoints for the construction of the spline, which is neither optimal nor valid for all values of $z$. A better method to select the endpoints may make use of the remainders associated with the perturbation and WKB series, which are yet to be calculated.

\section*{Acknowledgments}

Over a decade and a half this research has involved several successive rounds of undergraduate and graduate students at Texas
A\&M: Jeff Bouas, Fernando Daniel Mera, Cynthia Trendafilova, Krishna Thapa, Colin Whisler, Steven Murray, Hamilton Carter, David
Lujan, in addition to the 
current coauthors. Likewise at Oklahoma, Kim Milton has led a large succession of postdocs
and students: Prachi Parashar, Pushpa Kalauni, Taylor Murphy, Li Yang, Alex Mau, Jacob Tice, Elom Abalo, Nima
Portolami. Special recognition goes to Jef Wagner, who was a graduate student at Oklahoma and briefly a postdoc
(Visiting Assistant Professor) at A\&M. Others who have contributed include Martin Schaden (deceased), Lev
Kaplan, Ricardo Estrada, Klaus Kirsten, and Alejandro Satz.

At various times, these projects received support from the National Science Foundation, the Julian Schwinger
Foundation, the Simons Foundation, Texas A\&M Physics Department student research support funds, and an NSF REU
grant to the University of Oklahoma. The new research\cite{Shayit} was supported by undergraduate research funds from the Texas A\&M University Department of Physics and Astronomy, and its numerical portion relied on the advanced computing resources provided by Texas A\&M High Performance Research Computing.

 Vivek Sarin, who served as co-advisor on \rr\citen{Shayit}, provided guidance  about the numerical solution scheme. Kimball Milton gave constructive feedback
 and provided the data from \rr\citen{interior}  used in Figs.~\ref{fig:T00quadraticwall} and \ref{fig:T00RelErrorsquadraticwall}. 
 
 David Lujan participated in some of the earliest ``new'' work on this project, and Timothy Bates participated in the most recent parts and provided valuable comments on the manuscript.
 %who explored an alternative to the asymptotic expansion of the Green function based on \citen{faierman}.
 Helpful remarks were made by Colin Whisler, Pushpa Kalauni, Gerard Kennedy,  Justin Wilson, and Nima Pourtolami.

\setcounter{table}{0}
\setcounter{figure}{0}
\renewcommand{\thetable}{\Alph{section}\arabic{table}}
\renewcommand{\thefigure}{\Alph{section}\arabic{figure}}

\appendix
\section{The Numerical Solution}\label{appendix:numericalApprox}

\subsection{Discretization}

In Secs.~\ref{section:thePressureInside} and \ref{section:theEnergyDensityInside}, we use the numerical counterpart of $\overline{I}$, namely $\underline{I}$, to generate numerical approximations of the energy density and pressure. As the analogue of the operator \eqref{MethodsApproxIop}, it is composed of the numerical Green function, which is related to the numerical basis solutions $\underline F_\kappa(z)$ and $\underline G_\kappa(z)$ via Eq.~\eqref{IntroInterior210}. To obtain them, we convert the differential equation \eqref{IntroODE} to a matrix equation. This discretization takes the boundary conditions \eqref{IntroBoundaryConds} into account, which results in the nonsingularity of the matrix. The discretized equation may be solved by consecutive Gaussian eliminations for all desired values of $\kappa$. In each iteration, $\kappa$ is treated as a constant parameter.
For numerical stability we used a Cholesky factorization.  Details are provided in \rr\citen{Shayit}.

We discretize the entities composing Eq.~\eqref{IntroODE} as
\begin{equation}\label{MethodsDiscretizedDomainODE}
   \vec{z} =         
    \begin{bmatrix}
    z_0 \\
    z_1 \\
    \vdots \\
    \vdots \\
    z_{m + 1}
    \end{bmatrix}
,\quad
  \vec{V} =         
    \begin{bmatrix}
    V\left( z_{0} \right) \\
    V\left( z_{1} \right) \\
    \vdots \\
    \vdots \\
    V\left( z_{m+1} \right)
    \end{bmatrix}
,\quad
    \vec{y} =         
    \begin{bmatrix}
    y_{0} \\
    y_{1} \\
    \vdots \\
    \vdots \\
    y_{m+1}
    \end{bmatrix},
\end{equation}
 where $V(z)$ is the soft wall \eqref{IntroSoftWallPotential}.
We denote the uniform interval width by $h \equiv z_{i+1} - z_i$.

Since the values $y_0$ and  $y_{m+1}$ are known at the boundary points $z_0, z_{m+1}$, we solve Eq.~\eqref{IntroODE} in the domain $\left[z_1, \cdots, z_m \right]$. With this goal in mind, we seek a discretization of the differential operator acting on the left hand side of Eq.~\eqref{IntroODE} which yields $m$ equations for the $m$ unknowns $\{y_1, \cdots, y_m \}$.
The second derivative is approximated by the sixth-order centered finite difference formula\cite{finiteDifferenceAlgorithm,finiteDifferenceCoeffs}
\begin{eqnarray}\label{MethodsFiniteDifferenceCentered}
    180 h^2 \partial_z^2 y \big |_{z_i} &\approx& 2y_{i-3 }-27y_{i-2} + 270 y_{i-1} -490 y_i \nonumber \\ &\quad& + 270 y_{i+1} -27 y_{i+2} + 2 y_{i+3}, 
\end{eqnarray}
for $2 < i < m-1$. Near the boundary, we use the formulae \cite{finiteDifferenceAlgorithm,finiteDifferenceCoeffs}
\begin{eqnarray}\label{MethodsFiniteDifferenceSided}
    180 h^2 \partial_z^2 y \big |_{z_1} &\approx& 126 y_0 -70 y_1 - 486 y_2 + 855 y_3 \nonumber \\ &\quad& - 670 y_4 + 324 y_5 - 90 y_6 + 11 y_7, \nonumber\\
    180 h^2 \partial_z^2 y \big |_{z_2} &\approx& -11 y_0 + 214 y_1 - 378 y_2 + 130 y_3 \nonumber \\ &\quad& + 85 y_4 -54 y_5 + 16 y_6 - 2 y_7, \nonumber\\
    180 h^2 \partial_z^2 y \big |_{z_{m-1}} &\approx& -2 y_{m-6} + 16 y_{m-5} - 54 y_{m-4} \nonumber \\ &\quad& + 85 y_{m-3} + 130 y_{m-2} - 378 y_{m-1} \nonumber \\ &\quad& + 214 y_{m} - 11 y_{m+1}, \nonumber\\
    180 h^2 \partial_z^2 y \big |_{z_m} &\approx& 11 y_{m-6} - 90y_{m-5} + 324y_{m-4} \nonumber \\ &\quad& - 670y_{m-3} + 855 y_{m-2} - 486 y_{m-1} \nonumber \\ &\quad& - 70 y_{m} + 126y_{m+1}.
\end{eqnarray}
The portion of the formulae independent of the boundary conditions $y_0$ and $y_{m+1} $ can be combined into an $m \times m$ matrix. This matrix will operate on the column vector $\left[y_1, \cdots, y_m \right]^T$ to produce $\partial_z^2 y$ inside the interval.

The boundary terms can be added to the appropriate elements after the multiplication. According to the formulae \eqref{MethodsFiniteDifferenceCentered} and \eqref{MethodsFiniteDifferenceSided}, the boundary terms appear in the approximations for $y''(z_1)$, $y''(z_2)$, $y''(z_3)$, $y''(z_{m-2})$, $y''(z_{m-1})$, and $y''(z_m)$. We then approximate $\partial_z^2 y$ inside the interval as
\begin{eqnarray}\label{MethodsD2Matrix}  
    &&\begin{bmatrix}
    y''(z_1) \\
    y''(z_2) \\
    y''(z_3) \\
    y''(z_4) \\
    y''(z_5) \\
    \vdots \\
    y''(z_{m-4}) \\
    y''(z_{m-3}) \\
    y''(z_{m-2}) \\
    y''(z_{m-1}) \\
    y''(z_m)
    \end{bmatrix}
    \approx \\ \nonumber &&\underbrace{
    \frac{1}{180 h^2}
    \begin{bmatrix}
    -70 & -486 & 855 & -670 & 324 & -90 & 11 & 0 & \cdots & 0 \\
    214 & -378 & 130 & 85 & -54 & 16 & -2 & 0 &\cdots & 0 \\
    -27 & 270 & -490 & 270 & -27 & 2 & 0 & 0 &  \cdots & 0 \\
    2 & -27 & 270 & -490 & 270 & -27 & 2 & 0 &\cdots & 0 \\
    0 & 2 & -27 & 270 & -490 & 270 & -27 & 2 & \cdots & \vdots \\
    \vdots &  & \ddots & \ddots & \ddots & \ddots & \ddots & \ddots & \ddots &  \vdots\\
    \vdots &  &  & \ddots & \ddots & \ddots & \ddots & \ddots & \ddots & 2\\
    0 & \cdots & 0 &0 & 2 & -27 & 270 & -490 & 270 & -27\\
    0 & \cdots & 0 & -2 & 16 & -54 & 85 & 130 & -378 & 214\\
    0 & \cdots  &0 & 11 & -90 & 324 & -670 & 855 & -486 & -70
    \end{bmatrix}}_{D^2_z}
    \begin{bmatrix}
    y_{1} \\
    y_{2} \\
    y_{3}\\
    y_{4}\\
    y_{5}\\
    \vdots \\
    y_{m-4}\\
    y_{m-3}\\
    y_{m-2}\\
    y_{m-1}\\
    y_{m}
    \end{bmatrix} 
   % \\ \nonumber     &&{}
   +    \underbrace{\frac{1}{180 h^2}
     \begin{bmatrix}
    126 y_0 \\
    -11y_0 \\
    2 y_0\\
    0 \\
    0 \\
    \vdots \\
    0 \\
    0 \\
    2 y_{m+1}\\
    -11y_{m+1} \\
    126 y_{m+1}
    \end{bmatrix}}_{\vec{b}}.
\end{eqnarray}
 Finally, using the discretization \eqref{MethodsD2Matrix}, we convert Eq.~\eqref{IntroODE} into the matrix equation
\begin{equation}\label{MethodsDiscretizedODE}
    \left(-D^2_z + \mathop{\mathrm{diag}}\! \left(\vec V\right) + \kappa^2 \mathbb{I} \right)  
    \begin{bmatrix}
    y_{1} \\
    \vdots \\
    y_{m}
    \end{bmatrix} = \vec{b}.
\end{equation}

%Since the rows of $\mathcal{L}_m$ correspond to conditions on the solution at distinct points, they are linearly independent, and $\mathcal{L}_m$ is nonsingular. Therefore, the boundary conditions $y_0$ and $y_{m+1}$ (which define $\vec b$) uniquely determine the solution of Eq.~\eqref{MethodsDiscretizedODE}, as intended.

\subsection{Boundary Conditions}

To solve for $F_\kappa(z)$, we approximate the decaying behavior by taking a sufficiently large value of $z_{m+1}$ and demanding that $F_\kappa(z_{m+1})$ effectively vanishes. Using Eq.~\eqref{MethodsWKBApproximations}, we conclude that $z_{m+1} = 10$ can be taken to be the interval endpoint for our purposes.  

Since both conditions for $G_\kappa(z)$ in Eq.~\eqref{IntroBoundaryConds} are given at $z=0$, we cannot solve for it using Eq.~\eqref{MethodsDiscretizedODE} directly. However, we can solve a related boundary value problem, obtain the solution $\widetilde {G}_\kappa(z)$, and deduce $G_\kappa(z)$ from it. Since $G_\kappa(z)$ is defined to be exponentially increasing, we enforce the boundary condition $\widetilde {G}_\kappa(0) = G_\kappa(0) = 0$, and guess a larger boundary condition for $\widetilde {G}_\kappa(z)$ at a reasonable endpoint $z_{m+1}$, as shown in Table~\ref{tab:MethodsNumericalBoundaryConds}. The function $\widetilde {G}_\kappa(z)/\widetilde {G}_\kappa'(0)$ clearly satisfies the initial conditions \eqref{IntroBoundaryConds} and must coincide with $G_\kappa(z)$ from uniqueness arguments.

\begin{table}[h!]
        \tbl{Summary of the boundary conditions used to solve Eq.~\eqref{MethodsDiscretizedODE}.}{
        \begin{tabular}{ccccc}\toprule
		{Function} & {$z_0$} & {$y_0$} & {$z_{m+1}$} & {$y_{m+1}$}\\ \colrule
		\newline$F_\kappa(z)$ & 0 & 1 & 10 & 0\\
		\newline$\widetilde{G}_\kappa(z)$ & 0 & 0 & 2 & 1\\\botrule
        \end{tabular}\label{tab:MethodsNumericalBoundaryConds}}
\end{table}

\setcounter{table}{0}
\setcounter{figure}{0}
\section{Preliminary Methodologies}
\label{appendix:methods}

The central problem in this work
is bridging the gap between perturbative and WKB approximations.
In the exterior of the soft wall, such an operation was needed
only once: for the function called $\gamma_-(\kappa)$ in
\rr\citen{swout} (Sec.~IV and Figs.\ 6--7).  A simple spline between
low-order expansions was remarkably successful there.  In
retrospect, this was a stroke of luck; in the interior
calculation, bridging was needed more often, and the elementary
methodology was much less successful.  Solving this problem took
five years and the participation of three ``generations" of
undergraduate research assistants.  One lesson is that no method
is guaranteed to be a permanent solution, and sometimes it pays
to return to methods previously abandoned.  Therefore, a brief
account of some false starts may be of value.

In our initial approach, we attempted to construct piecewise solutions for the various relevant functions, such as $F_\kappa(z), G_\kappa(z)$, and $c_F(\kappa)$ (the link between the WKB approximation of $F_\kappa(z)$ and the
initial data \eqref{IntroBoundaryConds}; see \rr\citen{swout}, (3.9a).

Initially we developed the
small-$\kappa$ and large-$\kappa$ expansions only to rather low order, with a minimum of computer assistance.  In particular, we
implemented \eqref{MethodsPerturbationCoefficients} only for $n=1$, because {\sl Mathematica\/}
cannot evaluate the integrals for $n=2$. We picked various simple spline forms to approximate the functions between these two regimes, and fit the parameters to best agree with both expansions. The partition of the $\kappa$ domain into the three regimes is described in Table~\ref{tab:HistoricalSummaryPiecewise}. The various spline forms tested are shown in Table~\ref{tab:HistoricalSummarySplineForms}. It rapidly became clear that no
single spline form could give acceptable results in all cases.

\begin{table}[h!]
        \tbl{The piecewise partition of the $\kappa$ domain used to construct the spline.}{
        \begin{tabular}{ccc}\toprule
		{Function} & {Description} & {Approximation type} \\\colrule
		\newline$l(\kappa)$ & the ``left'' expression & perturbation expansion \\
		\newline$r(\kappa)$ & the ``right'' expression & WKB expansion  \\
		\newline$m(\kappa)$ & the ``middle'' expression & two-parameter spline \\\botrule
        \end{tabular}\label{tab:HistoricalSummaryPiecewise}}
\end{table}

\begin{table}[h!]
    \tbl{The various spline forms tested.}{
    \begin{tabular}{cc}\toprule
	{Spline form} & {Formula} \\\colrule
	\newline linear & $m(\kappa) = A\kappa +B$ \\
	\newline quadratic & $m(\kappa) = A\kappa^2 + B$ \\
	\newline reciprocal & $m(\kappa) = (A\kappa + B)^{-1}$ \\
	\newline exponential & $m(\kappa) = \exp\left(A\kappa + B\right)$ \\\botrule
    \end{tabular}\label{tab:HistoricalSummarySplineForms}}
\end{table}
%The various splines were generated and compared to their numerical counterpart to judge their validity. 

Given a form of $m(\kappa)$, we searched for endpoints $\kappa_L$ and $\kappa_R$ which would mark the transitions between the functions in Table~\ref{tab:HistoricalSummaryPiecewise}. The conditions imposed on the spline demanded that $m$ matched $l$ and $r$ at $\kappa_L$ and $\kappa_R$, respectively, to first order. This amounted to finding a zero of the vector-valued function
\begin{equation*}
    \vec g(\vec x) = \begin{pmatrix}
    l'(x_1) - m'(x_1)\\
    r'(x_2) - m'(x_2)
    \end{pmatrix}.
\end{equation*}
The parameters $A$ and $B$ were implicitly functions of $x_1$ and $x_2$ as well, as the spline function $m$ needed to satisfy the conditions. Thus generically, both components of $\vec g$ were functions of both $x_1$ and $x_2$.

We then used a multidimensional version of Newton's method to iteratively find the zero of the function. Using the Jacobian matrix, $J_{ij} = \frac{\partial g_i}{\partial x_j}$, each iteration of the algorithm computed
\begin{equation*}
    \vec{x}_{n+1} = \vec{x}_n - J^{-1} \cdot \vec{g}(\vec{x}_n).
\end{equation*}
If the algorithm converged, the resulting values of $\kappa_L$, $\kappa_R$, $A$, and $B$ would give a spline approximation to the solution in the intermediate regime. 

Unfortunately, even when a
suitable spline was found and the Newton iteration converged, the
accuracy of the results often was disappointing.  We made some
strategic changes: first, to interpolate the Green function
$g_\kappa$ directly, instead of its ingredients $F_\kappa$ and
$G_\kappa$, and second, to replace the spline by a Pad\'e
approximant. The two-point $[\ell/m]$ Pad\'e approximant of a function $f(\kappa)$, as described in \rr\citen{cizek}, is the rational function $P_\ell(\kappa)/Q_m(\kappa)$ whose power series matches $f(\kappa)$ as much as possible at two given points. As shown in Sec.~IV of \rr\citen{cizek}, the matching requirements yield linear constraints on the coefficients of $P_\ell(\kappa)$ and $Q_m(\kappa)$. These equations are solved to determine the approximant $P_\ell(\kappa)/Q_m(\kappa)$.

We constructed the two-point $[4/4]$ Pad\'e approximant of $g_\kappa(z)$ at $\kappa=0$ and $\infty$, matching to the coefficients of the first four terms in \eqref{MethodsPerturbativeGreen} and
the first five in the expansion of the dominant term in \eqref{MethodsWKBGreen} as
a series in $1/\kappa$. The resulting approximant was differentiated twice with respect to $z$ to approximate $\partial_z ^2 g_\kappa(z)$. 

Initial results of the Pad\'e method
were encouraging, but again not sufficiently accurate for our
purposes.  When the order was increased, spurious poles appeared
in the approximant.  (This is a well known disease of the Pad\'e
method.)  A pole can be bypassed by constructing a spline across
it, but this extra \emph{ad hoc} step has no strong advantage
over a pure spline.

Meanwhile, numerical calculations were being carried out for comparison with all these analytical approximations. The basis functions $F\kappa(z)$ and
$G_\kappa(z)$ were computed by the {\tt NDSOLVE} routine in {\sl Mathematica\/}, as well as the Numerical Calculus package (to compute the solution's derivatives). The initial conditions for $F_\kappa(z)$ were obtained from its first-order Fr\"{o}man approximation (\rr\citen{interior}, Appendix~A) at $z = 10$. While this process often yielded a solution, it was unstable and failed unexpectedly for some values of $z$.

For both the numerical and the analytical
solutions, therefore, we abandoned our original preference for
low-tech methods.  The expansions \eqref{MethodsPerturbativeGreen} and \eqref{MethodsWKBGreen} have been
implemented to quite high orders and successfully bridged by
splines, as reported in Sec.~\ref{section:theIntermediateRegime}. The numerical work has been
moved from {\sl Mathematica\/} to professional
numerical-analysis software as reported in Sec.~\ref{section:theIntermediateRegime} and \ref{appendix:numericalApprox}.

\end{document}